\documentclass[conference]{IEEEtran}%% Packages / general settings
\usepackage{graphicx}
\usepackage{amsmath,amssymb,dsfont,bbm,epstopdf,pgfplots,mathtools,enumitem,mathrsfs, bbm, algpseudocode , graphicx, dblfloatfix} 
\usepackage[normalem]{ulem}
\usepackage{color}
\usepackage{cite}%hyperref}
\usepackage{graphics}
\usepackage{tikz}
\usepackage{pgfplots}
\usepackage{subcaption}
\usetikzlibrary{shapes, arrows, decorations.markings, arrows.meta}
\usetikzlibrary{patterns} % LATEX and plain TEX when using Tik Z
\allowdisplaybreaks
\graphicspath{{.}{./Figures/}} % path to figures
\allowdisplaybreaks[4] % Allows the align environment to wrap over pages.

\newtheorem{lemma}{Lemma}

\newtheorem{theorem}{Theorem}

\newcommand{\Tu}{{\mathcal K_{\text{U}}} }

\renewcommand{\Pr}{{\mathbb{P}}}

\newcommand{\vect}[1]{\boldsymbol{#1}}

\definecolor{ForestGreen}{rgb}{0.0, 0.5, 0.0}

\newcommand{\mw}[1]{{\color{black}#1}}

\renewcommand{\P}{\mathsf{P}}

\renewcommand{\P}{\mathsf{P}}

\newcommand{\U}{\mathsf{U}}
\newcommand{\e}{\mathsf{e}}
\newcommand{\bu}{\beta_{\U}}
\newcommand{\bef}{\beta_{\e,1}}
\newcommand{\bes}{\beta_{\e,2}}
\newcommand{\be}{\beta_{\e}}
\newcommand{\xkef}{\vect X_{k}^{(\e,1)}}
\newcommand{\xkes}{\vect X_{k}^{(\e,2)}}
\newcommand{\xku}{\vect X_{k}^{(\U)}}
\newcommand{\Nu}{n_{\U}}
\newcommand{\Ne}{n_{\e}}
\newcommand{\MkF}{M_k^{(\U)}}
\newcommand{\MkS}{M_k^{(\e)}}
\newcommand{\syv}{ \sigma_{y|v}^2}
\newcommand{\sy}{ \sigma_{1}^2}
\newcommand{\vnorm}{r_k}
\newcommand{\syf}{\sigma_{2}^2}
\newcommand{\sys}{\sigma_{3}^2}
\newcommand{\syfx}{\tilde \sigma_{2}^2}
\newcommand{\sysx}{\tilde \sigma_{3}^2}
\newcommand{\syst}{\sigma_{4}^2}
\newcommand{\syfxt}{\tilde \sigma_{1}^2}
\newcommand{\sysxt}{\tilde \sigma_{4}^2}
\newcommand{\akf}{\alpha_{k,1}}
\newcommand{\aks}{\alpha_{k,2}}
\newcommand{\hk}{h_{k,k}}
\newcommand{\hkk}{h_{k-1,k}}
\newcommand{\cf}{c_1}
\newcommand{\cs}{c_2}
\newcommand{\ct}{c_3}
\newcommand{\cff}{c_4}
\newcommand{\sz}{\sigma_{z}^2}
\newcommand{\szf}{\sigma_{z_1}^2}
\newcommand{\szs}{\sigma_{z_2}^2}
\newcommand{\szt}{\sigma_{z_3}^2}
\begin{document}
\title{Joint Coding of URLLC and eMBB in Wyner's Soft-Handoff Network in the Finite Blocklength Regime}
\author{\IEEEauthorblockN{Homa Nikbakht$^{1}$, Mich\`ele Wigger$^2$, Shlomo Shamai (Shitz)$^3$, Jean-Marie Gorce$^1$, and H.~Vincent Poor$^4$}
	\IEEEauthorblockA{$^1$CITI Laboratory, INRIA,   $\quad ^2$LTCI,   T$\acute{\mbox{e}}$l$\acute{\mbox{e}}$com Paris, IP Paris,  $\quad ^3$Technion, $\quad^4$ Princeton University \\
%	91120 Palaiseau,
		\{homa.nikbakht, jean-marie.gorce\}@inria.fr, michele.wigger@telecom-paris.fr,  \\ sshlomo@ee.technion.ac.il, poor@princeton.edu}}
%	\and
%	\IEEEauthorblockN{Shlomo Shamai (Shitz)}
%	\IEEEauthorblockA{Technion, 
%		sshlomo@ee.technion.ac.il}}
\maketitle

 \begin{abstract}
Wyner's soft-handoff network is considered where transmitters simultaneously send messages of enhanced mobile broadband (eMBB) and ultra-reliable low-latency communication (URLLC) services. Due to the low-latency requirements, the URLLC messages are transmitted over fewer  channel uses compared  to the eMBB messages. To improve the reliability of the URLLC transmissions, we propose a coding scheme with finite blocklength codewords that exploits  dirty-paper coding (DPC) to precancel the interference from eMBB transmissions. Rigorous bounds are derived for the error probabilities of eMBB and URLLC transmissions achieved by our scheme. Numerical results illustrate that they are lower than for standard time-sharing. 
\end{abstract}

\section{Introduction}
The fifth  and the forthcoming sixth generations of mobile communications have to accommodate both ultra-reliable and low-latency communication (URLLC) and enhanced mobile broadband (eMBB) services \cite{Tataria201216g,Popovski2019}. URLLC services aim at guaranteeing high-reliability at a maximum end-to-end delay of $1$ms and are used for delay-sensitive applications such as  industrial control management as well as autonomous vehicle and remote surgery applications \cite{Popovski2019}. On the other hand,  eMBB services aim to provide high data rates and are used for delay-tolerant applications such as video streaming, virtual and augmented reality applications \cite{Bairagi, Anand2020}.  

The difference in the latency requirements of eMBB and URLLC services  along with the fact that that they are scheduled in the same frequency band make their coexistence challenging. Networks with such mixed-delay constraints have been studied recently. See \cite{HomaEntropy2022,Tajer2021,shlomo2012ISIT, HomaITW2020, Cohen2022} for a comprehensive review on related works. The previous studies are mostly focused on the performance of such networks in the asymptotic regime where the number of channel uses goes to infinity. Since the URLLC delay constraint limits the number of available channel uses, the problem of  joint coding of messages with heterogeneous blocklengths is of an increasing interest. %new characterizations of fundamental tradeoffs between the size of the message set, the probability of error, and the length of the code with finite blocklength codewords of both URLLC and eMBB transmissions  are of a growing interest. %The performance of such networks is mainly studied in the asymptotic information theory regime where the number of channel uses goes to infinity.  In the finite blocklength regime, 
%the work in \cite{shlomo2012ISIT} proposes a broadcast approach for a point-to-point scenario where the transmitter  manages to send URLLC messages over single coherence blocks of the fading channel,  and  simultaneously also sends eMBB messages over multiple coherence blocks.  
Notably, for the Gaussian point-to-point channel with messages of heterogeneous decoding deadlines, the work in \cite{ISIT2022long}, proposes a coding scheme which  exploits  dirty-paper coding (DPC) \cite{Costa1983, Scarlett2015}. Accounting for finite decoding deadline constraints,  rigorous bounds are derived on the achievable error probabilities of the messages. Their numerical results illustrate that their proposed scheme outperforms time sharing for a wide range of blocklengths.
% In a point-to-point setup, the work in~\cite{Cohen2022} proposes a novel layering scheme for data streaming under mixed-delay constraints where the base layer contains URLLC data and the enhancement layer  contains eMBB data.  In the proposed scheme, the base layer is  encoded  using a broadcast approach, which allows the receiver to  decode the URLLC data with minimum delay required.
 For the Gaussian broadcast channel with heterogeneous blocklength constraints, the work in \cite{Lin2021}, proposes a coding scheme which decodes the messages at time-instances that  depend on the realizations of the random channel fading. The authors showed that significant improvements are possible over standard successive interference cancellation. {In \cite{Mross2022}  achievable rates and  latency of the early-decoding scheme  in \cite{Lin2021} are improved by introducing \emph{concatenated shell codes}. Finally, \cite{Kassab2018} and \cite{HomaITW2019} studied the uplink of the cloud radio access networks where URLLC  messages are directly decoded at the base stations whereas  decoding of eMBB messages can be delayed to the cloud center. In particular, \cite{Kassab2018} performs a hybrid analysis where URLLC transmissions are studied in the finite blocklength regime and eMBB transmissions in the asymptotic infinite blocklength regime.} %They analyze the achievable rate of the URLLC transmissions in the finite block length regime.  }

In this paper, we consider  Wyner's soft-handoff model with $K$ interfering transmitters and receivers pairs. Each transmitter wishes to simultaneously transmit two messages of heterogeneous blocklengths: an URLLC message and an eMBB message. The URLLC message is transmitted over a shorter blocklength compared to the eMBB message.  Txs can hold a conferencing communication that depends only on the eMBB messages but not on the URLLC messages. By exploiting the DPC principle in \cite{Costa1983, Caire2003}, we propose a coding scheme to jointly transmit the URLLC and eMBB messages. Unlike \cite{ISIT2022long,Lin2021,Kassab2018}, we consider that codebooks  are generated randomly according to independent uniform distributions on the power-shell. Rigorous bounds are derived for achievable error probabilities of eMBB and URLLC transmissions.
To this end,   Gel'fand-Pinsker analysis techniques for finite blocklengths in \cite{Scarlett2015} are combined with the  multiple parallel channels approach in \cite{Erseghe2016}. Numerical results illustrate that our proposed scheme significantly outperforms  standard time-sharing. %The enhancement layer is encoded using a priori and posteriori forward error correction  so as to  be able to control the throughput-delay trade-off of this communication. % thus exploiting the ergodic behaviour of the channel.

 %eMBB and URLLC services are scheduled  in the same frequency band where their different delay requirements make their coexistence challenging. 

\section{Problem Setup}\label{sec:DescriptionOfTheProblem}
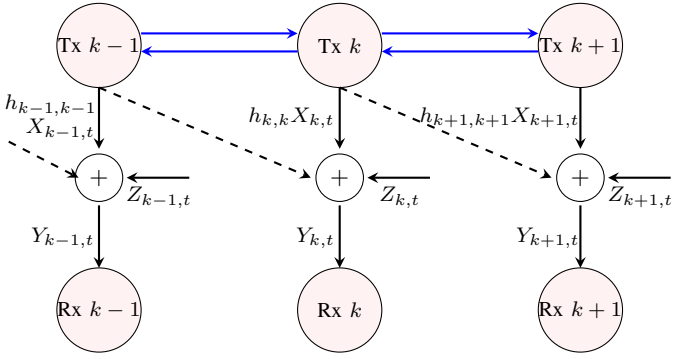
\begin{figure}[t]
%\small
  \centering
  \small
    %\hspace*{32pt}
 \begin{tikzpicture}[scale=1.6, >=stealth]
\centering
\tikzstyle{every node}=[draw,shape=circle, node distance=1cm];
\draw [fill = pink!20] (0,2.35) circle (0.35);
\draw [fill = pink!20] (0,0.15) circle (0.35);
\draw [fill = pink!20] (2,2.35) circle (0.35);
\draw [fill = pink!20] (2,0.15) circle (0.35);
\draw [fill = pink!20] (4,2.35) circle (0.35);
\draw [fill = pink!20] (4,0.15) circle (0.35);
%\draw (-0.5,0) -- (0.5,0) -- (0.5,0.5) -- (-0.5,0.5) -- (-0.5,0);
%\draw (-0.5,2) -- (0.5,2) -- (0.5,2.5) -- (-0.5,2.5) -- (-0.5,2);
%\draw (1.5,2) -- (2.5,2) -- (2.5,2.5) -- (1.5,2.5) -- (1.5,2);
%\draw (1.5,0) -- (2.5,0) -- (2.5,0.5) -- (1.5,0.5) -- (1.5,0);
%\draw (3.5,2) -- (4.5,2) -- (4.5,2.5) -- (3.5,2.5) -- (3.5,2);
%\draw (3.5,0) -- (4.5,0) -- (4.5,0.5) -- (3.5,0.5) -- (3.5,0);
\node [draw] at (0,1.25) {$+$};
\node [draw] at (2,1.25) {$+$};
\node [draw] at (4,1.25) {$+$};
\draw   [thick,->] (0,2)--(0,1.49);
\draw   [thick,->] (2,2)--(2,1.49);
\draw   [thick,->] (4,2)--(4,1.49);
\draw   [thick,->] (0,1.02)--(0,0.5);
\draw   [thick,->] (2,1.02)--(2,0.5);
\draw   [thick,->] (4,1.02)--(4,0.5);
\draw   [thick,->] (0.75,1.25)--(0.23,1.25);
\draw   [thick,->] (2.75,1.25)--(2.23,1.25);
\draw   [thick,->] (4.75,1.25)--(4.23,1.25);
\draw   [thick,->,dashed] (0,2)--(1.76,1.25);
\draw   [thick,->,dashed] (2,2)--(3.76,1.25);
\draw   [thick,->,dashed] (-0.75,1.55)--(-0.17,1.25);
%\draw   [thick,->,blue] (0.35,0.25)--(1.65,0.25);
%\draw   [thick,->,blue] (1.65,0.1)--(0.35,0.1);
%\draw   [thick,->,blue] (2.35,0.25)--(3.65,0.25);
%\draw   [thick,->,blue] (3.65,0.1)--(2.35,0.1);
%\draw   [thick,->,blue] (4.35,0.25)--(5,0.25);
%\draw   [thick,->,blue] (5,0.1)--(4.35,0.1);
%\draw   [thick,->,blue] (-1,0.25)--(-0.35,0.25);
%\draw   [thick,->,blue] (-0.35,0.1)--(-1,0.1);
%
\draw   [thick,->,blue] (0.35,0.35+2.1)--(1.65,0.35+2.1);
\draw   [thick,->,blue] (1.65,0.1+2.2)--(0.35,0.1+2.2);
\draw   [thick,->,blue] (2.35,0.35+2.1)--(3.65,0.35+2.1);
\draw   [thick,->,blue] (3.65,0.1+2.2)--(2.35,0.1+2.2);
%\draw   [thick,->,blue] (4.35,0.35+2.1)--(5,0.35+2.1);
%\draw   [thick,->,blue] (5,0.1+2.2)--(4.35,0.1+2.2);
%\draw   [thick,->,blue] (-1,0.35+2.1)--(-0.35,0.35+2.1);
%\draw   [thick,->,blue] (-0.35,0.1+2.2)--(-1,0.1+2.2);
%%%%%%%
\node [draw=none] at (-0.4,1.85) {\footnotesize{$h_{k-1,k-1}$}};
\node [draw=none] at (-0.32,1.75-0.1) {\footnotesize{$ X_{k-1,t}$}};
\node [draw=none] at (1.8-0.2,1.75) {\footnotesize$\hk  X_{k,t}$};
\node [draw=none] at (3.73-0.4,1.75) {\footnotesize$h_{k+1,k+1} X_{k+1,t}$};
\node [draw=none] at (0.5,1.39-0.3) {\footnotesize$Z_{k-1,t}$};
\node [draw=none] at (2.5,1.39-0.3) {\footnotesize$Z_{k,t}$};
\node [draw=none] at (4.5,1.39-0.3) {\footnotesize $Z_{k+1,t}$};
\node [draw=none] at (-0.3,0.75) {\footnotesize$Y_{k-1,t}$};
\node [draw=none] at (1.8,0.75) {\footnotesize$ Y_{k,t}$};
\node [draw=none] at (3.73,0.75) {\footnotesize$Y_{k+1,t}$};
\node [draw=none] at (0,0.15) {\footnotesize Rx~$k-1$};
\node [draw=none] at (2,0.15) {\footnotesize Rx $k$};
\node [draw=none] at (4,0.15) {\footnotesize Rx~$k+1$};
\node [draw=none] at (0,2.35) {\footnotesize Tx~$k-1$};
\node [draw=none] at (2,2.35) {\footnotesize Tx $k$};
\node [draw=none] at (4,2.35) {\footnotesize Tx~$k+1$};
%\node [draw=none] at (1.5,1.75-0.4) {\footnotesize$\hkk X_{k-1,t}$};
%\node [draw=none] at (0,2.9) {\small{$\{M_{k-1}^{(F)},M_{k-1}^{(S)}\}$}};
%\node [draw=none] at (2,2.9) {\small{$\{M_{k}^{(F)},M_{k}^{(S)}\}$}};
%\node [draw=none] at (4,2.9) {\small{$ \{M_{k+1}^{(F)},M_{k+1}^{(S)}\}$}};
%\node [draw=none,blue] at (1,0.45) {\small{$Q^{(j)}_{k-1 \rightarrow k}$}};
%\node [draw=none,blue] at (1,-0.1) {\small{$Q^{(j)}_{k \rightarrow k-1}$}};
%\node [draw=none,blue] at (3,0.45) {\small{$Q^{(j)}_{k \rightarrow k+1}$}};
%\node [draw=none,blue] at (3,-0.1) {\small{$Q^{(j)}_{k+1 \rightarrow k}$}};
%
%\node [draw=none,blue] at (1,2.65) {\footnotesize{$T^{(j)}_{k-1 \rightarrow k}$}};
%\node [draw=none,blue] at (1,2.1) {\footnotesize{$T^{(j)}_{k \rightarrow k-1}$}};
%\node [draw=none,blue] at (3,2.65) {\footnotesize{$T^{(j)}_{k \rightarrow k+1}$}};
%\node [draw=none,blue] at (3,2.1) {\footnotesize{$T^{(j)}_{k+1 \rightarrow k}$}};
%\node [draw=none] at (3,0.4) {$\hat{{M}}_{k}^{(F)}$};
%\node [draw=none, rotate = 90] (v1) at (1.58,0.4) {$\ldots$};
%\draw   [thick,latex'-latex'] (1.5,1.1)-- node [draw=none,  text width=3.0cm, shape=rectangle, pos=0.5, yshift=0.2cm,xshift=0.6cm ,midway, fill=none, node distance=1cm] {\small{$K$\text{subfiles}}}  (3,1.1);
%\draw   [thick,latex'-latex'] (1.4,1)-- node [draw=none,  text width=1.7cm, shape=rectangle, pos=0.5, yshift=3,xshift=-5 ,midway, fill=none, node distance=1cm,rotate = 90] {\small{$m\;$\text{files}}}  (1.4,0);
%\node [draw=none] at (2.25,-0.2) {$\tilde{L}$};
\end{tikzpicture}
%\vspace*{-5ex}

  \caption{System model. }
  \label{fig1}
%  \vspace*{-2ex}
\end{figure}~~%

Consider Wyner's soft-handoff network with $K$ transmitters (Txs) and ${K}$ receivers (Rxs)  that are aligned on two parallel lines so that each Tx $k$  has two neighbours, Tx~$k-1$ and Tx~$k+1$, and each Rx~$k$ has two neighbours, Rx~$k-1$ and Rx~$k+1$. Define $\mathcal K := \{1, \ldots, K\}$. The signal transmitted by Tx~$k \in \mathcal K$ is observed by Rx~$k$ and the neighboring Rx~$k+1$. See Figure~\ref{fig1}. 
Each Tx~$k \in \mathcal K$ sends a so called \emph{eMBB} type message $\MkS$ to its corresponding Rx~$k$, for $\MkS$  uniformly distributed over $\mathcal{M}^{(\e)}_k := \{1, \ldots, L_{\e}\}$.  {A subset of Txs $\mathcal{K}_{\U}\subset \mathcal{K}$ also sends additional \emph{URLLC} messages $\MkF$, for $k\in \mathcal{K}_{\U}$, for $\MkF$  uniformly distributed over the set $\mathcal{M}^{(\U)}:= \{1, \ldots, L_{\U}\}$. 
We  assume that 
\begin{equation}
\Tu := \{1,3, \ldots, K-1\},
\end{equation}
so that    URLLC transmissions are only interfered by the eMBB transmissions but  not by  other URLLC transmissions.  (The study of  sets $\mathcal{K}_{\U}$  with interfering URLLC messages is left as a future research direction.)
}

Communication takes place in two phases.

\noindent\underline{\textit{Tx-cooperation phase:}} \\
{The encoding  starts with a first \emph{Tx-cooperation phase} in which   Txs  share their eMBB message with their neighbouring Txs in $\Tu$.  (For example over high-rate optical fibers if the Txs are BSs.)} % over orthogonal backhaul links by exchanging a single message related to the eMBB messages in the system.  For tractability, this communication is assumed to be noise-free and of unlimited rate. Let $T_{\ell\to k}$ denote the cooperation message sent from a Tx~$\ell$ to its neighbouring Tx k. Notice that 
%\begin{equation}
%T_{\ell \to k}=\psi_{k\to \ell}\left(M_{\ell}^{(\e)}\right)
%\end{equation} for an appropriate encoding function $\psi_{\ell\to k}$.
The  URLLC messages, which are subject to stringent delay constraints, are only  generated after the Tx-cooperation phase, at the beginning of the subsequent   \emph{channel transmission phase}. 

\noindent\underline{\textit{Channel transmission phase:}}\\
URLLC messages are transmitted over $n_\U$ channel uses and eMBB messages over $\Ne>\Nu$ channel uses. The blocklengths $\Nu$ and $\Ne$ are assumed to be fixed constants. Notice that while the transmission delay of URLLC messages is determined by the $\Nu$ channel uses, transmission delay of eMBB consists  of both  the delay of the Tx-cooperation phase as well as the delay induced by the $\Ne$ channel uses.

For each $k \in \mathcal K$, Tx~$k$ computes its time-$t$ channel input $X_{k,t}$  with $t \in \{1, \ldots, n_e\}$ as 
\begin{IEEEeqnarray}{rCl}
X_{k,t} = \begin{cases}f_k^{(b)} \big( \MkF, \MkS,  T_{\ell \to k} \big), & k \in \Tu \; \; \text{and} \; \; t \le \Nu \\
f_k^{(e)} \big( \MkS,  T_{\ell \to k}\big), &  \hspace{-0.55cm}k \notin \Tu \; \; \text{or}\; \; \Nu < t \le \Ne,
\end{cases} \nonumber
\end{IEEEeqnarray}
for each $\ell \in \{k-1, k+1\} $ and for some encoding functions $f_k^{(b)}$ and  $f_k^{(e)}$ on appropriate domains satisfying 
the average block-power constraint
\begin{equation}\label{eq:power}
\frac{1}{\Ne} \sum_{t=1}^{\Ne} X_{k,t}^2
\leq \P, 
\quad \forall\ k \in \mathcal K, \qquad \textnormal{almost surely.}
\end{equation}
%Given the structure of the encoding functions,  observations of each Rx~$k \in \Tu$ can be viewed as arising from two parallel channels: over the first channel of $\Nu$ blocks both $\MkF$ and $\MkS$ are \emph{jointly} transmitted; over the second channel of $n_e - \Nu$ blocks only  $\MkS$ is transmitted. Our goal is to establish bounds on the average transmission rate of eMBB messages while keeping the error probability of the transmission of each URLLC message below $\epsilon_u$. %when it is assumed that the symbols arising from the second message are Gaussian. 

%Define the following channel input vectors
%\begin{subequations}
%	\begin{IEEEeqnarray}{rCl}
%		\mathbf X_1 &:=& \{X_{a_1}, \ldots, X_{a_2-1}\}, \\
%		\mathbf X_2&:=& \{X_{a_2}, \ldots, X_{d_{k,1}}\}, \\
%		\mathbf X_3&:=& \{X_{d_{k,1}+1}, \ldots, X_{d_{k,2}}\}. 
%	\end{IEEEeqnarray}
%\end{subequations}

The input-output relation of the network is  described as
\begin{equation}\label{Eqn:Channel}
{Y}_{k,t} = h_{k,k} {X}_{k,t} +h_{k-1,k} {X}_{ k-1,t}+ {Z}_{k,t},
\end{equation}
where $\{Z_{k,t}\}$ are independent and identically distributed (i.i.d.) standard Gaussian for all $k$ and $t$ and independent of all messages; $h_{k,\ell} > 0$ is the fixed channel coefficient between Tx~$k$ and Rx~$\ell$;   and we define $X_{0,t} = 0$ for all $t$.

After $\Nu$ channel uses, each  Rx~$k \in \Tu$ decodes the URLLC message $\MkF$ based on its own channel outputs $\vect Y_k^{\Nu} := \{Y_{k,1}, \ldots, Y_{k,\Nu}\}$. So, it produces:
\begin{equation}
	 \hat M_k^{(\U)}={g_k^{(\Nu)}}\big( \vect Y_k^{\Nu}\big),
	\end{equation} 
for some decoding function $g_k^{(\Nu)}$ on appropriate domains. The average error probability for each message $\MkF$ is given by
\begin{equation}
\epsilon_{\U,k} := \Pr\left \{ \hat M_k^{(\U)} \neq \MkF \right\}, \quad \text{for} \quad k \in \Tu.
\end{equation}
After $\Ne$ channel uses,  
each Rx~$k$ decodes its desired eMBB messages as
\begin{equation}\label{mhats}
\hat{{M}}_{k}^{(\e)}={b_{k}^{(\Ne)}}\left ( \vect Y_{k}^{\Ne} \right ),
\end{equation}
 where  $b_{k}^{(\Ne)}$ is a decoding function on appropriate domains. The average error probability for  message $\MkS$ is given by
\begin{equation}
\epsilon_{\e,k} := \Pr\left \{ \hat M_k^{(\e)} \neq \MkS\right\}, \quad \text{for} \quad k \in \mathcal K.
\end{equation}

	{We will be interested in the average URLLC  and eMBB error probabilities
\begin{IEEEeqnarray}{rCl}
 \epsilon_{\U}& := & \frac{1}{K} \sum_{k \in \Tu}   \epsilon_{\U,k},\\
 \epsilon_{\e}& :=& \frac{1}{K} \sum_{k \in \mathcal K}   \epsilon_{\e,k}.
\end{IEEEeqnarray}
}
%\Pr \bigg[\bigcup_{k \in \Tu} \!\!\big( \hat{M}_k^{(F)} \neq M_k^{(F)}\big)  \;\text{or} \! \!\bigcup_{k \in \Ta} \big(\hat{M}_k^{(S)} \neq M_k^{(S)}\big)  \bigg]  %\quad \textnormal{as } \quad n\to \infty.
%\end{equation}

\section{Coding Scheme}\label{coding}
%We employ time-sharing for the transmission of the URLLC messages at interfering Txs. (The scheme should thus perform well at reasonably high powers $P$ where time-sharing outperforms simultaneous transmission and treating interference as noise.) During the first $n_\U/2$ channel uses, only Txs with odd indices send their URLLC messages and during the subsequent $n_{\U}/2$ channel uses only Txs with even indices send URLLC messages. The URLLC blocklength $n_{\U}$ is thus assumed even. We only describe transmission in the first $n_{\U'}:=n_{\U}/2$ channel uses which we combine with the last $(\Ne-\Nu)/2$ channel uses where only eMBB messages are transmitted. 

%We pick the URLLC set $\Tu$ as 

%During the only Tx-cooperation round, each Tx in $\mathcal K \backslash \Tu$ shares its channel inputs, which it can precompute in advance since they only depend on an eMBB message, to the neighboring transmitter to its right.
{Txs in $\Tu$ use DPC  
 to precancel the interference of eMBB transmissions  from their neighbouring transmissions and from their own eMBB transmissions on their URLLC transmissions. (Recall that during the Tx-cooperation rounds   Txs in $\Tu$ learn the eMBB messages of their neighbouring Txs.)}

\subsection{Encoding at Txs in $\mathcal K \backslash \Tu$}
Each Tx~$k \in \mathcal K \backslash \Tu$ transmits only the eMBB message $\MkS$ over the entire block of $\Ne$ channel uses. %We divide $\Ne$ into two blocks of $\Nu$ and $\Ne - \Nu$ channel uses. Tx~$k$ transmits its eMBB message $\MkS$ over both blocks. The reason for this partitioning is that due to the transmitted signals by Tx~$k-1$, which belongs to $\Tu$, Rx~$k$ observes different interfering signals at each block.
 Over the first $\Nu$ channel uses, it transmits a codeword $\xkef(\MkS)$ that is uniformly distributed  on the centered $\Nu$-dimensional sphere of radius  $\sqrt{\Nu \beta_{\e} \P}$, for some $\beta_\e\in[0,1]$, independently of all other codewords. %, i.e., it is of density
%\begin{IEEEeqnarray}{rCl}\label{eq:fu}
%f_{\xkef} (\vect x_{k}^{(\e,1)}) = \frac{\delta \left (\left|\vect x_{k}^{(\e,1)}\righ|^2 - \Nu \be \P \right )}{S_{\Nu} (\sqrt{ \Nu \be \P })},
%\end{IEEEeqnarray}
%  where $\delta(\cdot)$ is the Dirac delta function, and 
%\begin{IEEEeqnarray}{rCl}
%S_{n}(r) = \frac{2\pi^{n/2}}{\Gamma(\frac{n}{2})}r^{n-1}
%\end{IEEEeqnarray}
%is the surface area of a sphere of radius $r$ in the $n$-dimensional Euclidean space.
Tx~$k$ also describes  its message $\MkS$, and thus its input signal $\xkef$, %and describes the quantized signal $\hat{\vect X}_k^{(\e,1)}$
 to the neigbouring Tx to its right during the only Tx-cooperation round. %(Recall that for simplicity, we assume   high cooperation rates.)% perfect quantization, meaning that $\hat{\vect X}_k^{(\e,1)}= {\vect X}_k^{(\e,1)}$.

 To encode $\MkS$ over the following $(\Ne - \Nu)$ channel uses, Tx~$k$ employs a second codeword $\xkes(\MkS)$ that is uniformly distributed  on the centered $(\Ne-\Nu)$-dimensional sphere of radius $\sqrt{(\Ne - \Nu) (1- \beta_{\e}) \P}$, independently of all other codewords.

  % over cooperation links.  

%In the following sections we provide  %  the interference of URLLC transmissions on Rxs observing only eMBB messages is 

\subsection{Encoding at Txs in $\Tu$}
Each Tx~$k \in \Tu$ has both eMBB and URLLC messages to transmit. 
%We denote by $\vect X_k^{(e)}$ the signal employed by Tx~$k$ to encode its eMBB message $\MkS$ and by $\vect X_k^{(u)}$ the signal employed by this Tx to encode its URLLC message $\MkF$. How this signals are formed is explained in the following subsections with details. 
To transmit its URLLC message $\MkF$, Tx~$k$ employs DPC encoding  to precancel the interference of the eMBB transmission of the Tx to its left and its own eMBB transmission.  %Assuming by $\hat{ \vect X}_{k-1,1}^{(e)}$ the signal it received from its neighbour Tx~$k-1$ and by $\vect X_{k,1}^{(e)}$ the signal that this Tx uses to encode its eMBB message $\MkS$. How this signals are formed is explained shortly in details. % We divide the total number of channel uses $n_e$  into $n_1$, $n_2$, $\tilde \Nu$ and $\tilde n_e$ blocks where 
%\begin{IEEEeqnarray}{rCl}
%\tilde \Nu = \Nu - n_1 - n_2, \quad \tilde n_e = n_e - \Nu 
%\end{IEEEeqnarray}
% Tx~$k$ then uses the first  $n_1$ channel uses to send the power type of the signal $\hat{ \vect X}_{k-1,3}^{(e)}$, and the following  $n_2$ channel uses to transmit the power type of its own signal $\vect X_{k,3}^{(e)}$.  % , then it sends its eMBB message $\MkS$ over all the remaining % As a result, in our scheme, we assign specific number of channel uses to transmit the power types of the interfering signals that we will explain in details shortly. %In our scheme, we assign $n_1$ channel uses to transmit the power type of the signal $\hat{\vect X}_{k-1}^{(n_e)}$ received from the Tx~$k-1$. To transmit the power type of the signal $\vect X_k^{(\Nu, e) }$ 
% We thus require to divide the total number of $n_e$ channel uses into four parts of $n_1$, $n_2$, $\Nu$ and $n_e - \Nu - n_1 - n_2$ channel uses. For simplicity, we call 
Tx~$k$ transmits its URLLC message over only $\Nu$ channel uses whereas it sends its eMBB message over the entire block of $\Ne$ channel uses. 
%\begin{equation}
%\tilde n_e = n_e - n_1 - n_2
%\end{equation} 
%channel uses whereas it sends  $\MkF$ over only $\Nu$ channel uses. Note that, we pick $n_1$ and $n_2$ such that $\Nu \le \tilde n_e$. 
To transmit both messages while satisfying \eqref{eq:power}, we divide the total transmit power $\P$ into three parts  $\bu \P$, $\bef  \P$, $\bes  \P$, where power $\bu\P$ is used for URLLC transmission, power $\bef\P$ for eMBB transmission during the first $\Nu$ channel uses, and power $\bes\P$ for eMBB transmission during the last $\Ne-\Nu$ channel uses. The coefficients $\bu$, $\bef$, $\bes \in [0,1]$ are chosen such that 
\begin{equation} \label{eq:10}
\bu + \bef + \bes = 1.
\end{equation}

\underline{\emph{Transmitting $\MkS$ and $\MkF$:}} 
Over the first $\Nu$ channel uses, Tx~$k$ sends its eMBB message $\MkS$ jointly with its URLLC message $\MkF$. To this end, it encodes $\MkS$ using a codeword $\xkef(\MkS)$ that is  uniformly distributed  on the centered $\Nu$-dimensional sphere of radius $\sqrt{\Nu \bef \P}$. To encode $\MkF$,  for each realization  $m$ of  message   $\MkF$, $\lfloor 2^{\Nu R_{\U}} \rfloor$ codewords $\vect V_{k}(m,i )$, $i=1,\ldots, \lfloor 2^{\Nu R_{\U}} \rfloor $,  are drawn uniformly from a centered $\Nu$-dimensional  sphere of radius $\sqrt{\vnorm \Nu \P}$ independently of each other and  of all other codewords, where
\begin{IEEEeqnarray}{rCl}
 \vnorm : = \bu + \alpha_{k,1}^2 \bef + \alpha_{k,2}^2 \be.
\end{IEEEeqnarray}
Tx $k$ then chooses a codeword   $\vect V_{k}(\MkF,i )$  such that the sequence 
\begin{equation} \label{eq:x21}
\xku :  = \vect V_{k}(\MkF,i )- \alpha_{k,1}  \xkef- \alpha_{k,2} {\vect X}_{k-1}^{(\e,1)}
\end{equation}
 lies in the set%satisfies $\xku \in \mathcal D_k$ where
\begin{IEEEeqnarray}{rCl}\label{eq:di}
\mathcal D_k := \left \{ \vect x_{k}^{(\U)} : \Nu \bu \P - \delta_{k} \le \left\|\vect x_{k}^{(\U)}\right\|^2 \le \Nu \bu \P \right \}
\end{IEEEeqnarray}
for a given $\delta_k> 0$. 
\mw{If multiple  such codewords  exist, one of them is chosen at random, and if no appropriate codeword exists, an error is declared.} %We denote this  error event by ${\mathcal E_{k,v}}$. 
%\begin{IEEEeqnarray}{rCl} \label{eq:ev}
%\lefteqn{\mathcal E_{k,v} :=  \Big \{\text{no}\; i \; \text{exists such that}} \notag \\
%&&  \hspace{0.95cm}  \; \vect V_k(\MkF, i) - \alpha_{k,1}  \xkef - \alpha_{k,2} {\vect X}_{k-1}^{(\e,1)} \in \mathcal D_k  \Big \}. \IEEEeqnarraynumspace
%\end{IEEEeqnarray} 
%

%All the auxiliary codebooks are revealed to all  Txs  and Rxs. {\color{blue}We will discuss the optimal choice of  $\alpha_{k,1}$ and $\alpha_{k,2}$.}

Over the first $\Nu$ channel uses, Tx~$k$ transmits
\begin{IEEEeqnarray}{rCl}
 \xku+ \xkef.
\end{IEEEeqnarray}

Over the last $(\Ne-\Nu)$ channel uses, Tx~$k$ simply encodes $\MkS$ using a codeword $\xkes(\MkS)$  that is uniformly distributed  on the centered $(\Ne - \Nu)$-dimensional sphere of radius $\sqrt{(\Ne - \Nu) \bes \P}$.

\subsection{Decoding at Rxs in $\mathcal K \backslash \Tu$}  
Each Rx~$k$ in $\mathcal K \backslash \Tu$ only has an eMBB message to decode. %Given the structure of the encoding functions, each
 Rx~$k \in \mathcal K \backslash \Tu$  decomposes  its channel outputs into two output blocks  consisting of the first  $\Nu$ and  the last $(\Ne - \Nu)$ channel uses, respectively. These blocks are of the form:% . The channel outputs are of the following form %: over the first channel of $n_1$ blocks the power type index 
\begin{subequations}
\begin{IEEEeqnarray}{rCl}
 \vect Y_{k,1} &=&  h_{k,k} \xkef + h_{k-1,k}( \vect X_{k-1}^{(\U)} + \vect X_{k-1}^{(\e,1)}) +  \vect Z_{k,1}, \IEEEeqnarraynumspace \\
\vect Y_{k,2} &=&  h_{k,k} \xkes + h_{k-1,k} \vect X_{k-1}^{(\e,2)}+ \vect Z_{k,2},
\end{IEEEeqnarray}
\end{subequations}
where $\vect Z_{k,1}$ and $\vect Z_{k,2}$  are   independent i.i.d. standard Gaussian noise sequences. 
For $ \vect Y_{k,1}=  \vect y_{k,1}$ and  $\vect Y_{k,2}=  \vect y_{k,2}$, Rx~$k$ estimates $\MkS$ as an index  $m$ for which the corresponding codewords  $\vect x_{k}^{(\e,1)}(m)$ and $\vect x_{k}^{(\e,2)}(m)$ maximize the information density 
\begin{IEEEeqnarray}{rCl}
\lefteqn{i_1(\vect x_{k}^{(\e,1)}, \vect x_{k}^{(\e,2)}; \vect y_{k,1}, \vect y_{k,2} )} \notag \\
&&:= \ln \frac{f_{\vect Y_{k,1} |\xkef } (\vect y_{k,1}| \vect  x_{k}^{(\e,1)} ) f_{\vect Y_{k,2}| \xkes } (\vect y_{k,2}| \vect x_{k}^{(\e,2)})}{f_{\vect Y_{k,1}} (\vect y_{k,1}) f_{\vect Y_{k,2}} (\vect y_{k,2})}, \IEEEeqnarraynumspace
\end{IEEEeqnarray}
among  all codeword pairs $\vect x_{k}^{(\e,1)}=\vect x_{k}^{(\e,1)}(m')$ and $\vect x_{k}^{(\e,2)}=\vect x_{k}^{(\e,2)} (m')$.
%Let $\mathcal E^{(\e)}_{k,1}$ denote the  error event  $\hat{M}_{k}^{(\e)} \neq  \MkS$.
%\begin{IEEEeqnarray}{rCl}\label{eq:ee1}
%\mathcal E^{(\e)}_{k,1} &:=&  \{ \text{Rx~$k \in \mathcal K \backslash \Tu$ chooses} \; \hat{M}_{k}^{(\e)} \neq  \MkS \}. \IEEEeqnarraynumspace
%\end{IEEEeqnarray}

 \subsection{Decoding at Rxs in $\Tu$}
Similarly to the previous subsection, also Rxs in $\Tu$  decompose  their channel outputs into two output blocks  consisting of the first  $\Nu$ and  the last $(\Ne - \Nu)$ channel uses, respectively. For a Rx~$k \in \Tu$, these blocks are of the form:
%in Given the structure of the encoding functions, each Rx~$k \in \Tu$ observes its channel outputs as arising from two channels: over the first channel of $\Nu$ blocks both $\MkF$ and $\MkS$ are transmitted; and over the second channel of $\Ne -  \Nu$ blocks only $\MkS$ is transmitted. Denoting by $\vect Y_{k,1}$ and $\vect Y_{k,2}$ as the channel outputs of the first and second channels, these outputs are given by
%\begin{subequations}
%\begin{IEEEeqnarray}{rCl}
% Y_{k,1, t} &=&  X_{k,t}^{(T,1)} + Z_{k,t},  \quad \text{for} \quad t \in \{1, \ldots, n_1\} \\
%Y_{k,2, t} &=& X_{k,t}^{(T,2)} + Z_{k,t},  \quad \text{for} \quad t \in \{n_1+1, \ldots, n_1+ n_2\} \\
%Y_{k,3,t} &=& U_{k,t} - \alpha_{k,1}  \vect X_{k,t}^{(e)} - \alpha_{k,2} \hat{\vect X}_{k-1,t}^{(e)} + \vect X_k^{(e)} + \vect X_{k-1}^{(e)}+ \vect Z_k
%\end{IEEEeqnarray}
%\end{subequations}
\begin{subequations}
\begin{IEEEeqnarray}{rCl}
\vect Y_{k,1} &=& h_{k,k} (\xku+\xkef )+h_{k-1,k} \vect X_{k-1}^{(\e,1)}+ \vect Z_{k,1}, \IEEEeqnarraynumspace \label{eq:y3}\\
\vect Y_{k,2} & = & h_{k,k}\xkes + h_{k-1,k}\vect X_{k-1}^{(\e,2)} + \vect Z_{k,2}.
\end{IEEEeqnarray}
\end{subequations}
where $\vect Z_{k,1}$ and $\vect Z_{k,2}$  are   independent  i.i.d. standard Gaussian noise sequences. 

\subsubsection{ Decoding $\MkF$}
 Rx~$k$ decodes $\MkF$ based on the outputs of  the first channel inputs $\vect Y_{k,1}$ defined in \eqref{eq:y3}. 
%\begin{equation} \label{eq:yk3}
%\vect Y_{k,3} = \vect U_k - \alpha_{k,1}  \vect X_k^{(e)} - \alpha_{k,2} \hat{\vect X}_{k-1}^{(e)} + \vect X_k^{(e)} + \vect X_{k-1}^{(e)}+ \vect Z_k
%\end{equation}
Rx~$k$  estimates $\MkF$ as an index  $m$ for which the corresponding codeword  $\vect v_{k}(m,i)$  maximizes the information density  
\begin{equation}
i(\vect v_k; \vect y_{k,1} ):= \ln \frac{f_{\vect Y_{k,1} | \vect V_k} (\vect y_{k,1}| \vect v_k) }{f_{\vect Y_{k,1}} (\vect y_{k,1})},
\end{equation} 
among all codewords $\vect v_k=\vect v_{k}(m',j)$. %Let $\mathcal{E}^{(\U)}_{k}$ denote the  error event  $\hat{M}_{k}^{(\U)} \neq  \MkF$.

%We have the following error event while decoding $\MkF$: 
%\begin{IEEEeqnarray}{rCl} \label{eq:eu}
%\mathcal E_{k}^{(\U)} &:=&  \left\{ \text{Rx~$k$ chooses} \; \hat{M}_k^{(\U)} \neq \MkF \right \}. 
%\end{IEEEeqnarray}

\subsubsection{Decoding $\MkS$} \label{sec:decembb}
 Rx~$k$ decodes $\MkS$ based on the channel outputs of the first and second channels $\vect Y_{k,1}$ and $\vect Y_{k,2}$ by looking for the index $m$ for which the corresponding codewords $\vect x_{k}^{(\e,1)}(m)$ and $\vect x_{k}^{(\e,2)}(m)$ maximize the information density 
\begin{IEEEeqnarray}{rCl}
\lefteqn{i_2(\vect x_{k}^{(\e,1)}, \vect x_{k}^{(\e,2)}; \vect y_{k,1}, \vect y_{k,2} )} \notag \\
&:=& \ln \frac{f_{\vect Y_{k,1} |\xkef } (\vect y_{k,1}| \vect  x_{k}^{(\e,1)} ) f_{\vect Y_{k,2}| \xkes } (\vect y_{k,2}| \vect x_{k}^{(\e,2)})}{f_{\vect Y_{k,1}} (\vect y_{k,1}) f_{\vect Y_{k,2}} (\vect y_{k,2})} \IEEEeqnarraynumspace
\end{IEEEeqnarray}
%over all $\vect x_{k}^{(\e,1)} \in \mathbb R^{\Nu}$ and $\vect x_{k}^{(\e,2)} \in \mathbb R^{\Ne - \Nu}$.  
 among  all codeword pairs $\vect x_{k}^{(\e,1)}(m')$ and $\vect x_{k}^{(\e,2)} (m')$. % for $m'\neq {m}$.
%Let $\mathcal E^{(\e)}_{k,1}$ denote the  error event  $\hat{M}_{k}^{(\e)} \neq  \MkS$.
\begin{figure*}[b!] 
\hrule
%\begin{subequations}\label{eq:37}
\begin{IEEEeqnarray}{rCl}
\mathcal L_{k,1} &:=& \frac{\delta_{k}}{2\Nu \P \sqrt{\pi \vnorm }}\frac{\Gamma(\frac{\Nu}{2})}{\Gamma (\frac{\Nu-1}{2})} \frac{1}{\akf \sqrt{\bef}} \left (1 -{\left (\vnorm - \bu + \aks \sqrt{\be}(\akf \sqrt{\bef} +  \sqrt{ \vnorm} )+ \frac{\delta_{k}}{2  \Nu \P } \right )^2}/ (\akf^2 \vnorm \bef) \right)^{\frac{\Nu-3}{2}}, \label{eq:L1} \\
\mathcal L_{k,2}& :=& \frac{\delta_{k}}{2\Nu \P \sqrt{\pi \vnorm }}\frac{\Gamma(\frac{\Nu}{2})}{\Gamma (\frac{\Nu-1}{2})} \frac{1}{\aks \sqrt{\be}} \left (1 -{\left (\vnorm - \bu + \akf \sqrt{\bef}(\aks \sqrt{\be} +  \sqrt{ \vnorm} )+ \frac{\delta_{k}}{2  \Nu \P } \right )^2}/(\aks^2 \vnorm \be) \right)^{\frac{\Nu-3}{2}}, \label{eq:L2}\\
\bar \gamma_{\e,1} & := &  \Nu \left (\ln(\syf) - \P  \left ( \left ( \cf + \hk \sqrt{\be} \right )^2 - \frac{\cf^2}{\syf}   \right ) \right )+ (\Ne - \Nu) \left ( \ln(\sys) - \P \left ( \left (\ct +  \hk \sqrt{ 1- \be}\right )^2 -  \frac{\ct^2}{\sys}   \right ) \right) -\frac{2\gamma_{\e,1} }{J_{\e,1}}, \label{eq:bgef} \\
\bar \gamma_{\e,2} & := & \Nu \left (\ln (\sy) -\P\left ( \left (\cs + \hk \sqrt{\bef}\right )^2 -\frac{\cs^2}{\sy} \right ) \right) + (\Ne - \Nu)  \left (\ln (\syst)- \P \left (\left (\cff +\hk \sqrt{\bes}\right )^2   -\frac{\cff^2}{\syst}  \right ) \right) -  \frac{2\gamma_{\e,2} }{J_{\e,2}}. \label{eq:bges}\IEEEeqnarraynumspace
\end{IEEEeqnarray}
%\end{subequations}
\end{figure*}
\section{Main Result}
Fix $\be ,\bef,\bes,\bu\in[0,1]$ such that \eqref{eq:10} is satisfied.
%\begin{IEEEeqnarray}{rCl}
%\bef + \bes + \bu = 1.
%\end{IEEEeqnarray}
Define
\begin{subequations}
\begin{IEEEeqnarray}{rCl}
\sy &:=& \hk^2 \left (\vnorm +(1 - \akf)^2  \bef \right) \P \notag \\
&& + (\hkk- \hk \aks)^2 \be \P+ 1,  \label{eq:s1}\\
\syf & := & \hkk^2 \left (\vnorm + ( 1- \alpha_{k-1})^2\bef \right ) \P \notag \\
&& +\hkk^2 \alpha_{k-1,2}^2 \be\P +  \hk^2 \be \P +1,  \label{eq:s2}\\
\sys & := & \left (h_{k,k}^2(1- \be) + h_{k-1,k}^2 \bes \right ) \P + 1, \label{eq:s3}\\
\notag \\
\syst & := & \left (h_{k,k}^2\bes + h_{k-1,k}^2 (1-\be) \right ) \P + 1, \label{eq:s4} \\
\cf &:=&  \left (h_{k,k} \sqrt{\be} + h_{k-1,k}(\sqrt{\bu}+ \sqrt{\bef} ) \right ), \label{eq:c1}\\
\cs & :=& \left (h_{k,k} (\sqrt{\bu} +\sqrt{\bef}) + h_{k-1,k} \sqrt{\be} \right ), \label{eq:c2} \\
\ct & := & \left (h_{k,k} \sqrt{1 - \be} + h_{k-1,k} \sqrt{\bes}\right ), \label{eq:c3}\\
\cff &:=& \left (h_{k,k} \sqrt{\bes} + h_{k-1,k} \sqrt{( 1- \be)}  \right ). \label{eq:c4}
\end{IEEEeqnarray}
\end{subequations}
By employing the scheme proposed in Section~\ref{coding}, we have the following theorem on the upper bounds on the average  \mw{URLLC and eMBB error probabilities $\epsilon_{\U}$ and  $\epsilon_{\e}$.}

\begin{theorem}\label{th1}
For fixed message set sizes $L_{\U}$ and $L_{\e}$, the average error probabilities $\epsilon_{\U}$ and $\epsilon_{\e}$ are bounded by 
\mw{\begin{IEEEeqnarray}{rCl}
\epsilon_{\U} & \le &\frac{1}{K} \sum_{k \in \Tu}  \Big (1- F \left (u_{k,2} - u_{k,1}\right ) + F \left (- u_{k,2} - u_{k,1} \right ) \notag \\
&+&  \left ( 1-  \max \{\mathcal L_{k,1}, \mathcal L_{k,2} \}\right )^{\lfloor 2^{\Nu R_v}\rfloor} \Big)+L_{\U} \lfloor 2^{\Nu R_v}\rfloor e^{-\gamma_{\U}}, \label{boundu} \IEEEeqnarraynumspace\\
\epsilon_{\e} & \le & \frac{1}{K} \sum_{k \in \mathcal K \backslash \Tu} \frac{1}{\bar \gamma_{\e,1}} \left ( \zeta_1 l_{k,1}  +\zeta_2 l_{k,2} +  \Nu l_{3} + (\Ne -\Nu) l_{4}  \right ) \notag \\
&+& \frac{1}{K} \sum_{k \in \Tu} \frac{1}{\bar \gamma_{\e,2}} \left ( \zeta_1 d_{k,1} + \zeta_2 d_{k,2} +  \Nu d_{3} + (\Ne -\Nu) d_{4} \right )\notag \\
&+&  L_{\e} (e^{-\gamma_{\e,1}} + e^{-\gamma_{\e,2}} ),\label{bounde}
\end{IEEEeqnarray}
for any $\gamma_{\U}$, $\gamma_{\e,1}$ and $\gamma_{\e,2}$, and where 
\begin{IEEEeqnarray}{rCl}
u_{k,1} & := &\sqrt{\Nu \P}\left ( \cs + \frac{\hk \sqrt{\vnorm} \sy}{\sy -1} \right), \\
u_{k,2} & := & \sqrt{\frac{\sy}{\sy -1} \left (  \Nu \ln (\sy) - \frac{2\gamma_1}{J_{\U}} + \frac{\Nu \P \vnorm \hk^2}{\sy} \right )},  \\
l_{k,1} & := &2 \sqrt{\Nu \P}  \left ( \cf + \hk \sqrt{\be} - \frac{\cf}{\syf}  \right ), \label{eq:lk1}  \\
l_{k,2} & := & 2\sqrt{(\Ne - \Nu) \P} \left ( \ct   + h_{k,k}\sqrt{1 - \be} - \frac{\ct}{\sys}\right ), \label{eq:lk2}\IEEEeqnarraynumspace \\
d_{k,1} & := &2 \sqrt{\Nu \P} \left (\cs + \hk \sqrt{\bef} - \frac{\cs}{\sy } \ \right ), \label{eq:dk1} \IEEEeqnarraynumspace  \\
d_{k,2} & := & 2\sqrt{ (\Ne - \Nu) \P} \left ( \cff  + h_{k,k} \sqrt{\bes} -\frac{\cff}{\syst} \right ), \label{eq:dk2}\IEEEeqnarraynumspace
\end{IEEEeqnarray}
and $l_{3} := (\syf - 1) /\syf$, $l_{4} := (\sys - 1)/\sys$, $d_{3} := (\sy - 1)/\sy$, $d_{4} := (\syst - 1)/\syst $,}  $ \zeta_1 := \sqrt{2} \Gamma(\frac{\Nu +1}{2})/ \Gamma(\frac{\Nu}{2})$, $\zeta_2 := \sqrt{2}\Gamma(\frac{\Ne- \Nu +1}{2}) / \Gamma(\frac{\Ne - \Nu}{2})$,  $\mathcal L_{k,1}$, $\mathcal L_{k,2}$, $\bar \gamma_{\e,1}$, and $\bar \gamma_{\e,2}$ are defined in \eqref{eq:L1} to \eqref{eq:bges},  $J_{\U}$, $J_{\e,1}$ and $J_{\e,2}$ are defined in \eqref{eq:ju}, \eqref{eq:je1} and \eqref{eq:je2}, and $ F (\cdot)$ represents the cumulative distribution function (CDF) of a chi distribution of degree $\Nu$. 
\end{theorem}
\begin{IEEEproof}
See Appendix~\ref{sec:proofth1}. % for the proof of the bound in \eqref{boundu} and \cite{Longversion} for the proof of the bound in \eqref{bounde}. 
\end{IEEEproof}

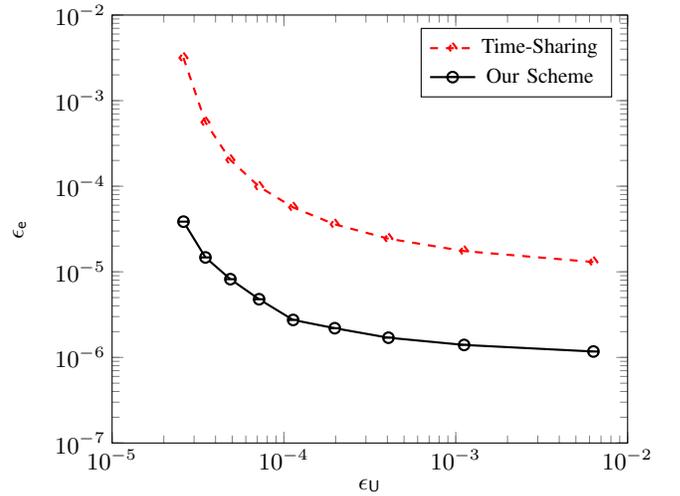
\begin{figure}[t!]
\centering
\begin{tikzpicture}[scale=1]
\begin{axis}[
    xlabel={\small {$\epsilon_{\U}$ }},
    ylabel={\small {$\epsilon_{\e}$ }},
     xlabel style={yshift=.5em},
     ylabel style={yshift=0em},
    xmin=1e-5, xmax=1e-2,
    ymin=1e-7, ymax=1e-2,
    xtick={1e-9,1e-8,1e-7,1e-6,1e-5,1e-4,1e-3,1e-2,1e-1,1},
    ytick={1e-9,1e-8,1e-7,1e-6,1e-5,1e-4,1e-3,1e-2,1e-1,1},
    yticklabel style = {font=\small,xshift=0.25ex},
    xticklabel style = {font=\small,yshift=0.25ex},
    legend pos=north east,
    ymode=log,
xmode = log,
]

% \addplot[ color=red, line width = 0.5mm] coordinates {  (2,0.999107822289119)(6,0.998778671116591)(10,0.994483015257301)(14,0.973986730045327)(18,0.878347923618621)(22,0.520354441213826)(26,0.034726808109444)(30,0.000000018585841)};

\addplot[ color=red, thick,  mark=diamond, dashed] coordinates { (0.00632455532033675,1.30134883134501e-05)(0.00111803398874990,1.74692810742171e-05)(0.000405720412966790,2.43924205986611e-05)(0.000197642353760524,3.58609569093279e-05)(0.000113137084989848,5.65685424949239e-05)(7.17219138186558e-05,9.88211768802618e-05)(4.87848411973222e-05,0.000202860206483395)(3.49385621484343e-05,0.000559016994374948)(2.60269766269002e-05,0.00316227766016837)};

\addplot[ color=black, thick,  mark=halfcircle] coordinates { (0.00632455532033675,1.17219591271469e-06)(0.00111803398874990,1.399107822289119e-6)(0.000405720412966790,1.698778671116591e-6)(0.000197642353760524,2.194483015257301e-6)(0.000113137084989848,0.273986730045327e-5)(7.17219138186558e-05,0.478347923618621e-05)(4.87848411973222e-05,0.00820354441213826e-3)(3.49385621484343e-05,0.014726808109444e-3)(2.60269766269002e-05,0.000038585841)};

\legend{ {\footnotesize Time-Sharing}, {\footnotesize Our Scheme}, {\footnotesize $\epsilon_{\e}, \Ne = 30$}, {\footnotesize $\epsilon_{\e}, \Ne = 36$}}  
\end{axis}

\vspace{-0.8cm}
\end{tikzpicture}

\caption{ $\epsilon_{\e}$ vs $\epsilon_{\U}$ for $\P = 10, \Ne = 100$ and $\Nu$ decreases from $90$ to $10$ with steps of $10$. }%, $\kappa = 1$, $\epsilon_T = 0.000001$ }
\label{fig2}
\vspace{-0.5cm}
\end{figure}

%\hn{
%\begin{remark}
%Under the assumption that $\Ne \to \infty$, the classical results are achievable for the eMBB transmissions, i.e., $\epsilon_{\e} = 0$. In this case, since $\Nu$ is negligible compared to $\Ne$, one can assume interference-free URLLC  transmissions over $\Nu$ channel by Txs in $\Tu$ which improves the bound in \eqref{boundu}. See \cite{Kassab2018} for the rate analysis of such a scenario.
%\end{remark}
%
%\begin{remark}
%To let all the Txs send URLLC messages, one can employ a time-sharing based scheme where during the first $\Nu/2$ channel uses, only Txs with odd indices send their URLLC messages and during the subsequent $\Nu/2$ channel uses only Txs with even indices send URLLC messages. Rxs then have to decompose their channel outputs into three output blocks consisting of the first and second $\Nu/2$ and the third $\Ne - \Nu$ channel uses. 
%\end{remark}
%}
In Figure~\ref{fig2}, we numerically compare the bounds in Theorem~\ref{th1} with the time-sharing scheme where only Txs in $\Tu$ send URLLC messages over $\Nu$ channel uses whereas all the Txs in $\mathcal K$ send eMBB messages but over only the remaining $\Ne - \Nu$ channel uses. In this plot, the value of $\Nu$ varies from $90$ to $10$ with  step size $10$, while the value of $\Ne$ is fixed at $100$. In our simulations, the values of the parameters $\be$, $\bu$, $\bef$, $\bes$, $\akf$ and $\aks$ are optimized to minimize $\epsilon_{\e}$ for a given $\epsilon_{\U}$.  As can be seem from this figure, our scheme outperforms the time-sharing scheme. 

\section{Conclusions}
{We considered Wyner's soft-handoff model where transmitters simultaneously  send eMBB and URLLC messages of heterogeneous blocklengths. We proposed a coding scheme to jointly transmit URLLC and eMBB messages in such a network. We derived rigorous upper bounds on the error probability of   eMBB and URLLC transmissions. Our numerical analysis showed that the proposed scheme significantly improves over  the standard time-sharing. 

An interesting future line of work is to study this network under the assumption that $\Ne$ is much larger than $\Nu$. This assumption allows the eMBB transmissions to benefit from their delay-tolerance feature. Another interesting scenario is to let all the Txs send URLLC messages which requires dealing with the interference from the URLLC messages on the URLLC transmissions as well.}

\section*{Acknowledgment}
The works of H. Nikbakht and J-M.~Gorce have been supported by INRIA Nokia Bell Labs ADR “Network Information Theory" 
and  by the French National Agency for Research (ANR) under grant ANR-16-CE25-0001 - ARBURST.
The work of M. Wigger has been supported by the
       European Union's Horizon 2020 Research and Innovation Program,
       grant agreements no. 715111. The work of S. Shamai (Shitz) has been supported
by the US-Israel Binational Science Foundation
(BSF) under grant BSF-2018710. The work of H. V. Poor
has been supported by the U.S. National Science Foundation (NSF) within the Israel-US Binational program
under grant CCF-1908308.
\appendices

\section{Proof of Theorem~\ref{th1}} \label{sec:proofth1}
\subsection{Bounding $\epsilon_{\U,k}$} 
We start by bounding the decoding error probability of  a  URLLC message at a given Rx~$k$ in $\Tu$. Define the decoding error event $\mathcal E_{k}^{(\U)}:=\{ \hat{M}_{k}^{(\U)} \neq  \MkF\}$ and let $\mathcal E_{k,v}$ be the encoding error event that no appropriate codeword $\vect{V}_k(\MkF,i)$ can be found so that $\vect{X}_k^{(\U)} (\MkF) \in \mathcal{D}_k$. We have:
\begin{IEEEeqnarray}{rCl}
\epsilon_{U,k}  \le \Pr [\mathcal E_{k,v}] +  \Pr [\mathcal E_{k}^{(\U)}| \mathcal E_{k,v}]. \IEEEeqnarraynumspace
\end{IEEEeqnarray}

\subsubsection{Analyzing $\Pr [\mathcal E_{k,v}]$}
To calculate this probability, we follow a similar argument as in \cite[Appendix E]{Scarlett2015}.  From \eqref{eq:di} we notice  that $(\vect V_k - \alpha_{k,1}  \xkef - \alpha_{k,2} {\vect X}_{k-1}^{(\e,1)} )\in \mathcal D_k$ if and only if 
\begin{IEEEeqnarray}{rCl}
 \Nu \bu \P - \delta_{k} \le ||\vect V_k - \alpha_{k,1}  \xkef - \alpha_{k,2} {\vect X}_{k-1}^{(\e,1)}||^2  \le \Nu \bu \P. \notag
\end{IEEEeqnarray}
Recall that  $||\vect V_k||^2  = \Nu \vnorm \P$ almost surely. Thus event $\mathcal{E}_{k,v}$ holds whenever the following condition is violated: 
\begin{IEEEeqnarray}{rCl} \label{eq:29}
\lefteqn{\Nu (\vnorm - \bu)\P +  ||\alpha_{k,1}  \xkef + \alpha_{k,2} \hat{\vect X}_{k-1}^{(\e,1)}||^2}\notag \\
 & \le& 2\alpha_{k,1} \langle\vect V_k,  \xkef \rangle +2 \alpha_{k,2} \langle \vect V_k ,  {\vect X}_{k-1}^{(\e,1)} \rangle  \notag \\
&\le&\Nu (\vnorm - \bu)\P +  ||\alpha_{k,1}  \xkef + \alpha_{k,2} {\vect X}_{k-1}^{(\e,1)}||^2 + \delta_{k}. \IEEEeqnarraynumspace
\end{IEEEeqnarray}
Define
\begin{IEEEeqnarray}{rCl}
C_k &:=& \frac{\Nu (\vnorm - \bu)\P}{2 \alpha_{k,1}} \notag \\
&+&  \frac{||\alpha_{k,1}  \xkef + \alpha_{k,2} {\vect X}_{k-1}^{(\e,1)}||^2}{2\alpha_{k,1}} - \frac{ \alpha_{k,2}}{\alpha_{k,1}} \langle \vect V_k ,  {\vect X}_{k-1}^{(\e,1)} \rangle. \IEEEeqnarraynumspace 
\end{IEEEeqnarray}
Equation \eqref{eq:29} then is equivalent to 
\begin{IEEEeqnarray}{rCl}\label{eq:Ck}
C_k \le \langle\vect V_k,  \xkef \rangle \le C_k + \frac{\delta_{k}}{2 \alpha_{k,1}}.
\end{IEEEeqnarray}

Since $ \xkef$ is drawn uniformly from the sphere, the distribution of  $\langle\vect V_k,   \xkef \rangle$  depends on $\vect V_k$ only through its magnitude, this is seen by noting that the inner product of two vectors is unchanged when an orthogonal  transformation is applied to both arguments, and the distribution of $ \xkef$ is unchanged under any orthogonal  transformation.  In the following we therefore assume that $\vect V_k = (||\vect V_k||, 0, \ldots, 0)$, in which case \eqref{eq:Ck} is equivalent to: 
\begin{IEEEeqnarray}{rCl}
\frac{C_k}{||\vect V_k||} \le X_{k,1}^{(\e,1)} \le \frac{C_k}{||\vect V_k||} + \frac{\delta_{k}}{2 \alpha_{k,1} ||\vect V_k||} 
\end{IEEEeqnarray}
where $X_{k,1}^{(\e,1)}$ is the first entry of the vector $\xkef$.
We conclude that $\Pr [\vect V_k - \alpha_{k,1}  \xkef - \alpha_{k,2} {\vect X}_{k-1}^{(\e,1)} \in \mathcal D_k]$ is lower bounded by the probability of the first entry of $\xkef$ falling into the  interval of length $\frac{\delta_{k}}{2 \alpha_{k,1} ||\vect V_k||} $ that starts at $\frac{C_k}{||\vect V_k||}$. (Notice that the length of the interval is deterministic because $\|\vect{V}_k\|=\sqrt{\Nu r_k \P}$ is  a constant, but its starting point is random because $C_k$ is a random variable. 

The distribution of a given symbol in a length-$\Nu$ random sequence distributed uniformly on the sphere is \cite{Stam1982}
\begin{IEEEeqnarray}{rCl}\label{eq:fx}
f_{X_{k,1}^{(\e,1)}}(x_{k,1}^{(\e,1)})&=& \frac{1}{\sqrt{\pi \Nu \bef \P}}\frac{\Gamma(\frac{\Nu}{2})}{\Gamma (\frac{\Nu-1}{2})} \left (1 - \frac{(x_{k,1}^{(\e,1)})^2}{\Nu \bef \P} \right)^{\frac{\Nu-3}{2}} \notag\\
&& \times \mathbbm{1}\{(x_{k,1}^{(\e,1)})^2 \le  \Nu \bef \P \}.
\end{IEEEeqnarray}
This density function is decreasing in $(x_{k,1}^{(\e,1)})^2$, which implies that
\begin{IEEEeqnarray}{rCl} \label{eq:pu}
\lefteqn{\Pr [\vect V_k - \alpha_{k,1}  \xkef - \alpha_{k,2} {\vect X}_{k-1}^{(\e,1)} \in \mathcal D_k] } \notag \\
&& \ge   \frac{\delta_{k}}{2 \alpha_{k,1} ||\vect V_k||}  f_{X_{k,1}^{(\e,1)}} \left ( \frac{C_k}{||\vect V_k||} + \frac{\delta_{k}}{2 \alpha_{k,1} ||\vect V_k||} \right).
\end{IEEEeqnarray}
Furthermore by the Cauchy-Schwartz inequality:
\begin{IEEEeqnarray}{rCl}
C_k &\le& \frac{\Nu (\alpha_{k,1}^2 \bef + \alpha_{k,2}^2 \be)\P}{2 \alpha_{k,1}} \notag \\
&& +  \frac{\alpha_{k,1}^2 || \xkef||^2 + \alpha_{k,2}^2 ||{\vect X}_{k-1}^{(\e,1)}||^2 }{2\alpha_{k,1}} \notag \\
&& + \frac{2 \akf \aks|| \xkef||\cdot ||{\vect X}_{k-1}^{(\e,1)}||}{2\alpha_{k,1}} \notag \\
&&  +  \frac{ \alpha_{k,2}}{\alpha_{k,1}} || \vect V_k|| \cdot ||{\vect X}_{k-1}^{(\e,1)}||.
%& = & \Nu \P \left ( \alpha_{k,1} \bef + \frac{\alpha_{k,2}^2 \be}{\alpha_{k,1}} + \aks \sqrt{\bef \be} + \frac{\alpha_{k,2}}{\alpha_{k,1}} \sqrt{\be \vnorm}\right )
\end{IEEEeqnarray}
Thus 
\begin{IEEEeqnarray}{rCl}
\frac{C_k}{||\vect V_k||} \le A_k,
\end{IEEEeqnarray}
where
\begin{IEEEeqnarray}{rCl}
A_k &:=&\sqrt{\Nu \P} \left ( \frac{\vnorm - \bu+ (\aks \sqrt{\be} + \akf \sqrt{\bef})^2 }{2\alpha_{k,1}\sqrt{\vnorm}} \right) \notag \\
 &&+ \frac{\aks \sqrt{\Nu \P \be}}{\akf}.
 \IEEEeqnarraynumspace
\end{IEEEeqnarray}
Therefore
\begin{IEEEeqnarray}{rCl} \label{eq:pu}
\lefteqn{\Pr [\vect V_k - \alpha_{k,1}  \xkef - \alpha_{k,2} \hat{\vect X}_{k-1}^{(\e,1)} \in \mathcal D_k] } \notag \\
&&\ge   \frac{\delta_{k}}{2 \alpha_{k,1} ||\vect V_k||}  f_{X_{k,1}^{(\e,1)}} \left ( A_k + \frac{\delta_{k}}{2 \alpha_{k,1} ||\vect V_k||} \right).
\end{IEEEeqnarray}

In a similar way,  in \eqref{eq:29}, one can move the term $2\alpha_{k,1} \langle\vect V_k,  \xkef \rangle$ to both sides and bound the above probability by the probability of the first entry of ${\vect X}_{k-1}^{(\e,1)}$ falling within a given interval. This leads to an equivalent bound, which combined with \eqref{eq:pu} yields:
\begin{IEEEeqnarray}{rCl}
\Pr [\vect V_k - \alpha_{k,1}  \xkef - \alpha_{k,2} {\vect X}_{k-1}^{(\e,1)} \in \mathcal D_k]  \ge \max \{\mathcal L_1, \mathcal L_2 \} ,\IEEEeqnarraynumspace
\end{IEEEeqnarray}
where $\mathcal L_1$ and $\mathcal L_2$ are defined in \eqref{eq:L1} and \eqref{eq:L2}, respectively. 
%\begin{IEEEeqnarray}{rCl}\label{eq:L}
%\mathcal L_1 &=&  \frac{\zeta}{\akf \sqrt{\bef}} \left (1 -\frac{\left (\vnorm - \bu + \aks \sqrt{\be}(\akf \sqrt{\bef} +  \sqrt{ \vnorm} )+ \frac{\delta_{k}}{2  \Nu \P } \right )^2}{\akf^2 \vnorm \bef} \right)^{\frac{\Nu-3}{2}} \\
%\mathcal L_2& =&  \frac{\zeta}{\aks \sqrt{\be}} \left (1 -\frac{\left (\vnorm - \bu + \akf \sqrt{\bef}(\aks \sqrt{\be} +  \sqrt{ \vnorm} )+ \frac{\delta_{k}}{2  \Nu \P } \right )^2}{\aks^2 \vnorm \be} \right)^{\frac{\Nu-3}{2}}
%\end{IEEEeqnarray}
%with 
%\begin{IEEEeqnarray}{rCl}
%\zeta = \frac{\delta_{k}}{2\Nu \P \sqrt{\pi \vnorm }}\frac{\Gamma(\frac{\Nu}{2})}{\Gamma (\frac{\Nu-1}{2})}
%\end{IEEEeqnarray}
Since the $\lfloor 2^{\Nu R_v}\rfloor$ codewords are generated independently, thus
\begin{IEEEeqnarray}{rCl}
\Pr [\mathcal E_{k,v}]
\le \left ( 1-  \max \{\mathcal L_1, \mathcal L_2 \}\right )^{\lfloor 2^{\Nu R_v}\rfloor}.
\end{IEEEeqnarray}

\subsubsection{Analyzing $ \Pr [\mathcal E_{k}^{(\U)}| \mathcal E_{k,v}] $} To evaluate this error event, we use the threshold bound for maximum-metric decoding. I.e.,
\begin{IEEEeqnarray}{rCl} \label{eq:43}
 \Pr [\mathcal E_{k}^{(\U)}| \mathcal E_{k,v}] &\le& \Pr[i(\vect V_k; \vect Y_{k,1} ) \le \gamma_{\U}]  \notag \\
&& + M_{\U} \lfloor 2^{\Nu R_v}\rfloor \cdot \mathbb P[i(\bar {\vect V}_k; \vect Y_{k,1} )> \gamma_{\U}] 
\end{IEEEeqnarray}
for any $\gamma_{\U}$, where $\bar{\vect V}_k \sim f_{\vect V_k}$ and is  independent of $(\vect V_k, \vect Y_{k,1})$. We start by calculating $\mathbb P[i(\bar{\vect V}_k, \vect Y_{k,1} )> \gamma_{\U}] $. By Bayes rule we have
\begin{IEEEeqnarray}{rCl}\label{eq:44}
f_{\vect V_k}(\bar {\vect v}_k) &=& \frac{f_{\vect Y_{k,1}}(\vect y_{k,1})f_{\vect V_k | \vect Y_{k,1}} (\bar {\vect v}_k | \vect y_{k,1}) }{f_{\vect Y_{k,1}| \vect V_k} (\vect y_{k,1}| \bar {\vect v}_k)}\\
& =& f_{\vect V_k | \vect Y_{k,1}} (\bar {\vect v}_k | \vect y_{k,1}) \exp \left ( - i(\bar{\vect v}_k, \vect y_{k,1} ) \right ). 
\end{IEEEeqnarray}
By multiplying both sides of the above equation by $\mathbbm {1} \{i(\bar{\vect v}_k, \vect y_{k,1} )> \gamma_{\U}\}$ and integrating over all $\bar{\vect v}_k$, we have
\begin{IEEEeqnarray}{rCl} \label{eq:45}
\lefteqn{\int_{\bar {\vect v}_k} \mathbbm {1} \{i(\bar{\vect v}_k, \vect y_{k,1} )> \gamma_{\U}\} f_{\vect V_k}(\bar {\vect v}_k) d \bar {\vect v}_k = } \notag \\
& &  \int_{\bar {\vect v}_k} \mathbbm {1} \{i(\bar{\vect v}_k, \vect y_{k,1} )> \gamma_{\U}\}  e^{  - i(\bar{\vect v}_k, \vect y_{k,1} )} f_{\vect V_k | \vect Y_{k,1}} (\bar {\vect v}_k | \vect y_{k,1}) d \bar {\vect v}_k. \IEEEeqnarraynumspace
\end{IEEEeqnarray}
Note that the left-hand side of \eqref{eq:45} is equivalent to $\Pr [i(\bar{\vect v}_k, \vect y_{k,1} )> \gamma_{\U} | \vect Y_{k,1} = \vect y_{k,1} ] $. Thus 
\begin{IEEEeqnarray}{rCl}
\lefteqn{\Pr [i(\bar{\vect v}_k, \vect y_{k,1} )> \gamma_{\U} | \vect Y_{k,1} = \vect y_{k,1} ] } \\
&= & \int_{\bar {\vect v}_k} \mathbbm {1} \{i(\bar{\vect v}_k, \vect y_{k,1} )> \gamma_{\U}\}  \notag \\
&& \hspace{0.5cm} \times \exp \left ( - i(\bar{\vect v}_k, \vect y_{k,1} ) \right ) f_{\vect V_k | \vect Y_{k,1}} (\bar {\vect v}_k | \vect y_{k,1}) d \bar {\vect v}_k \\
%& = & \int_{\bar {\vect v}_k} \mathbbm {1} \{\frac{f_{\vect Y_{k,1}| \vect V_k} (\vect y_{k,1}| \bar {\vect v}_k)}{f_{\vect Y_{k,1}}(\vect y_{k,1})}> e^{\gamma_{\U}}\}  \exp \left ( - i(\bar{\vect v}_k, \vect y_{k,1} ) \right ) f_{\vect V_k | \vect Y_{k,1}} (\bar {\vect v}_k | \vect y_{k,1}) d \bar {\vect v}_k \\
& = & \int_{\bar {\vect v}_k} \mathbbm {1} \{\frac{f_{\vect Y_{k,1}| \vect V_k} (\vect y_{k,1}| \bar {\vect v}_k)}{f_{\vect Y_{k,1}}(\vect y_{k,1})} e^{-\gamma_{\U}}>1\}  \notag \\
&& \hspace{0.5cm} \times \exp \left ( - i(\bar{\vect v}_k, \vect y_{k,1} ) \right ) f_{\vect V_k | \vect Y_{k,1}} (\bar {\vect v}_k | \vect y_{k,1}) d \bar {\vect v}_k \\
& \le & \int_{\bar {\vect v}_k} \frac{f_{\vect Y_{k,1}| \vect V_k}  (\vect y_{k,1}| \bar {\vect v}_k)}{f_{\vect Y_{k,1}}(\vect y_{k,1})} e^{-\gamma_{\U}}  \notag \\
&& \hspace{0.5cm} \times \exp \left ( - i(\bar{\vect v}_k, \vect y_{k,1} ) \right ) f_{\vect V_k | \vect Y_{k,1}} (\bar {\vect v}_k | \vect y_{k,1}) d \bar {\vect v}_k \\
& = & \int_{\bar {\vect v}_k} e^{-\gamma_{\U}} f_{\vect V_k | \vect Y_{k,1}} (\bar {\vect v}_k | \vect y_{k,1}) d \bar {\vect v}_k \\
& \le &  e^{-\gamma_{\U}}. \label{eq:52}
\end{IEEEeqnarray}
Now we calculate $\Pr[i(\vect V_k, \vect Y_{k,1} ) \le \gamma_{\U}] $. 
Note that $\vect Y_{k,1}$  and $\vect Y_{k,1}| \vect V_k$ do not follow a Gaussian distribution. 
Now define $Q^{(\U)} (\vect y_{k,1}) = \mathcal N(\vect y_{k,1}; \vect 0, I_n \sy)$ and $ W^{(\U)} (\vect y_{k,1}| \vect v_k) = \mathcal N(\vect y_{k,1}; h_{k,k} \vect V_k, I_n \syv)$ where $\sy$ is defined in \eqref{eq:s1} and $\syv =  1$.
%Also define the following typical output set 
%\begin{IEEEeqnarray}{rCl}
%\mathcal F_{k,1} = \{ \vect y_{k,1}: ||\vect y_{k,1}||^2 \in [\sy - \delta_{k,1}, \sy + \delta_{k,1}] \}
%\end{IEEEeqnarray}

Introduce 
\begin{IEEEeqnarray}{rCl}
\tilde i(\vect v_k; \vect y_{k,1} ) := \ln \frac{W^{(\U)}(\vect y_{k,1}| \vect v_k)}{Q^{(\U)}(\vect y_{k,1}) }.
\end{IEEEeqnarray}
\begin{lemma} \label{lemma1}
We can prove that 
\begin{IEEEeqnarray}{rCl}
 \frac{ i(\vect v_k; \vect y_{k,1} )}{\tilde i(\vect v_k; \vect y_{k,1} )} \ge J_{\U},
\end{IEEEeqnarray}
where
\begin{IEEEeqnarray}{rCl} \label{eq:ju}
J_{\U} & :=& (\Nu -2) \ln (2a_1a_2) \notag \\
&& - 2\Nu\P (a_1^2 \bef  + a_2^2 \be ) - \frac{e^{c_{\Gamma}} a_2^2 \be \P }{\sqrt{2\pi a_1^2 \bef \P} } - \kappa
\end{IEEEeqnarray}
and $a_1 := \hk(1 - \akf)$, $a_2 := \hkk - \hk \aks$,  $\kappa := \ln (\frac{1}{2} ) + c_{\Gamma} + \ln (\sqrt{\frac{\pi}{8}}) - 2 \ln (\hk)$ with $c_{\Gamma} \le 2$. 
\end{lemma}
\begin{IEEEproof}
We use Lemma~\ref{lemma7} of Appendix~\ref{sec:lemmas} to upper bound $f_{\vect Y_{k,1}}(\vect y_{k,1})  / Q^{(\U)} (\vect y_{k,1}) $ and Lemma~\ref{lemma6} of Appendix~\ref{sec:lemmas} to lower bound $f_{\vect Y_{k,1}| \vect V_k} (\vect y_{k,1}| {\vect v}_k) / W^{(\U)}(\vect y_{k,1}| \vect v_k)$. 
\end{IEEEproof}
As a result, we have 
\begin{IEEEeqnarray}{rCl}
  \lefteqn{  \Pr [i(\vect V_k; \vect Y_{k,1} ) \le \gamma_{\U} )]} \\
& \le & \Pr [\tilde i(\vect V_k; \vect Y_{k,1} ) \le \frac{\gamma_{\U} }{J_{\U}}] \\
& = &    \Pr  \left [\ln \frac{W^{(\U)}(\vect Y_{k,1}| \vect V_k)}{Q^{(\U)}(\vect Y_{k,1}) }  \le \frac{\gamma_{\U} }{J_{\U}} \right ] \\
& = & \Pr \Bigg [ \ln {\frac{\frac{1}{(\sqrt{2\syv\pi})^{\Nu}}\exp \left (- \frac{|| \vect Y_{k,1} - h_{k,k}\vect V_k||^2}{2\syv}\right )}{\frac{1}{(\sqrt{2\pi \sy})^{\Nu}}\exp \left (- \frac{|| \vect Y_{k,1}||^2}{2 \sy}\right )}} \le \frac{\gamma_{\U} }{J_{\U}}\Bigg ] \\
%&= & \Pr \Bigg [ \frac{\Nu}{2} \ln \frac{\sy}{\syv} + \frac{|| \vect Y_{k,1}||^2}{2 \sy} \notag \\
%&& \hspace{0.5cm}- \frac{|| \vect Y_{k,1} - h_{k,k}\vect V_k||^2}{2\syv} \le \frac{\gamma_{\U} }{J_{\U}}\Bigg ] \\
%& = & \Pr \Bigg [\left (\frac{1}{\sy}- \frac{1}{\syv}\right)|| h_{k,k} \xku + h_{k,k} \xkef + h_{k-1,k} {\vect X}_{k-1}^{(\e,1)}  + \vect Z_{k,1}||^2 \notag \\
%&& \hspace{2cm}-  \frac{h_{k,k}^2}{\syv}||\vect V_k||^2 + \frac{2h_{k,k}}{\syv} \langle \vect Y_{k,1}, \vect V_k \rangle \le \frac{2\gamma_{\U} }{J_{\U}} - \Nu\ln \frac{\sy}{\syv} \Bigg ] \IEEEeqnarraynumspace \\
%& = & \Pr \Bigg [\left (\frac{\sy - \syv}{\sy \syv}\right)|| h_{k,k} \xku + h_{k,k} \xkef + h_{k-1,k} {\vect X}_{k-1}^{(\e,1)}  + \vect Z_{k,1}||^2 \notag \\
%&& \hspace{2cm}+  \frac{h_{k,k}^2}{\syv}||\vect V_k||^2 - \frac{2h_{k,k}}{\syv} \langle \vect Y_{k,1}, \vect V_k \rangle \ge -\frac{2\gamma_{\U} }{J_{\U}} + \Nu\ln \frac{\sy}{\syv} \Bigg ] \IEEEeqnarraynumspace \\
%& = & \Pr \Bigg [ || h_{k,k} \xku + h_{k,k} \xkef + h_{k-1,k} {\vect X}_{k-1}^{(e,1)}  + \vect Z_{k,1}||^2  \notag \\
%&& - \frac{2\hk \sy}{\sy - \syv} \langle h_{k,k} \xku + h_{k,k} \xkef + h_{k-1,k} {\vect X}_{k-1}^{(e,1)}  + \vect Z_{k,1}, \vect V_k \rangle  \ge \tilde \gamma_{\U}\Bigg ]\\
& = & \Pr \Bigg [ h_{k,k}^2||\xku||^2 + h_{k,k}^2 ||\xkef||^2 + h_{k-1,k}^2 ||\vect X_{k-1}^{(\e,1)}||^2   \notag \\
&& \hspace{0.5cm} + ||\vect Z_{k,1}||^2+ 2 h_{k,k}^2\langle \xku, \xkef  \rangle \notag \\
&& \hspace{0.5cm} + 2 h_{k,k}h_{k-1,k} \langle \xku, \vect X_{k-1}^{(\e,1)}  \rangle  + 2 h_{k,k} \langle \xku, \vect Z_{k,1} \rangle\notag \\
&& \hspace{0.5cm}  + 2 \hk \hkk \langle \xkef,  \vect X_{k-1}^{(\e,1)}  \rangle + 2 \hk \langle \xkef, \vect Z_{k,1} \rangle \notag \\
&& \hspace{0.5cm} + 2 \hkk \langle \vect X_{k-1}^{(\e,1)} , \vect Z_{k,1} \rangle - \frac{2 \hk^2\sy}{ \sy -\syv}  \langle \xku, \vect V_k\rangle   \notag \\
&&\hspace{0.5cm} - \frac{2 \hk ^2\sy}{\sy - \syv}  \langle \xkef, \vect V_k\rangle  - \frac{2\hk \sy}{\sy - \syv} \langle \vect Z_{k,1}, \vect V_k\rangle  \notag \\
&&  \hspace{2.3cm}- \frac{2\hkk \hk \sy}{\sy -\syv} \langle \vect X_{k-1}^{(\e,1)}, \vect V_k\rangle \ge \tilde \gamma_{\U}\Bigg ] \\
& \le & \Pr \Bigg [ h_{k,k}^2\Nu \bu \P+ ||\vect Z_{k,1}||^2  \notag \\
&& \hspace{0.5cm}+ 2 h_{k,k}^2\Nu \P \sqrt{\bu \bef} + 2 h_{k,k}h_{k-1,k} \Nu \P \sqrt{\bu \be}   \notag \\
&& \hspace{0.5cm}+ 2 h_{k,k} \sqrt{\Nu \P \bu} \cdot || \vect Z_{k,1}|| + 2 \hk \hkk \Nu \P \sqrt{\bef \be}  \notag \\
&& \hspace{0.5cm}+ 2 \hk \sqrt{\Nu \P \bef} || \vect Z_{k,1}|| + 2 \hkk \sqrt{\Nu \P \be} || \vect Z_{k,1} || \notag \\
&& \hspace{0.5cm}+ \frac{2 \hk^2\sy}{ \sy - \syv}  \Nu \P \sqrt{\bu \vnorm}  + \frac{2 \hk^2 \sy}{\sy - \syv}  \Nu \P \sqrt{\bef \vnorm}  \notag \\
&&\hspace{0.5cm} + \frac{2 \hk \sy}{\sy - \syv} \left (\hkk \Nu \P \sqrt{\be \vnorm}  +\sqrt{\Nu \P \vnorm} || \vect Z_{k,1}|| \right) \notag \\
&& \hspace{1.5cm} \ge \tilde \gamma_{\U} - \Nu \P (h_{k,k}^2 \bef + h_{k-1,k}^2 \be)\Bigg ] \\
%& = & \Pr \Bigg [ ||\vect Z_{k,1}||^2 + c_1 ||\vect Z_{k,1}|| \ge  \tilde \gamma_{\U}- \Nu \P (h_{k,k}^2 \bef + h_{k-1,k}^2 \be) - \Nu \P \hk^2 \bu \notag \\
%&& \hspace{0.5cm} - 2\Nu \P \left ( \hk^2 \sqrt{\bu \bef} + \hk \hkk \sqrt{\be} \left (\sqrt{\bu} + \sqrt{\bef} \right )\right ) \notag \\
%&&\hspace{0.6cm}- \frac{2 \hk \sy \Nu \P \sqrt{\vnorm}}{\sy - \syv} \left ( \hk \sqrt{\bu} + \hk \sqrt{\bef} + \hkk \sqrt{\be} \right) \Bigg ] \\
& = & \Pr \left  [ ||\vect Z_{k,1}||^2 + b_1 ||\vect Z_{k,1}|| \ge \bar \gamma_{\U}\right ] \\
& = & \Pr \left  [ \left (||\vect Z_{k,1}|| + \frac{b_1}{2} \right ) ^2  \ge \bar \gamma_{\U} + \frac{b_1^2}{4}\right ] \\
%& = & 1- \Pr \left  [ - \sqrt{\bar \gamma_{\U} + \frac{c_1^2}{4}} - \frac{c_1}{2} \le  ||\vect Z_{k,1}|| \le \sqrt{\bar \gamma_{\U} + \frac{c_1^2}{4}} - \frac{c_1}{2} \right ] \\
& = & 1- F\left (\sqrt{\bar \gamma_{\U} + \frac{b_1^2}{4}} - \frac{b_1}{2}\right ) + F \left (-\sqrt{\bar \gamma_{\U} + \frac{b_1^2}{4}} - \frac{b_1}{2} \right )
%& = & 1- F_{||\vect Z_{k,1}||} \left (u_1- u_2\right ) + F_{||\vect Z_{k,1}||} \left (- u_1- u_2  \right )
\end{IEEEeqnarray}
where 
\begin{subequations}
\begin{IEEEeqnarray}{rCl}
 %c_1 &:=& 2\sqrt{\Nu \P}\left ( \hk \left ( \sqrt{\bu} + \sqrt{\bef} \right ) + \hkk \sqrt{\be} + \frac{\hk \sy \sqrt{\vnorm} }{\sy - 1} \right) \\
\tilde \gamma_{\U} & := &   \frac{\sy}{\sy - 1} \left ( \Nu \ln(\sy) -\frac{2\gamma_1}{J_{\U}} - \hk^2 \vnorm \Nu \P  \right),  \IEEEeqnarraynumspace \\
\bar \gamma_{\U}&:=& \tilde \gamma_{\U}  -{\Nu \P \cs^2}- \frac{2 \hk \sy \Nu \P \sqrt{\vnorm}}{\sy - 1}  \cs, \\
b_1 & := & 2\sqrt{\Nu \P}\left ( \cs + \frac{\hk \sqrt{\vnorm} \sy}{\sy -1} \right),  
%u_2 & = &\sqrt{\Nu \P}\left ( \hk \left ( \sqrt{\bu} + \sqrt{\bef} \right ) + \hkk \sqrt{\be} + \frac{\hk \sqrt{\vnorm} \sy}{\sy -1} \right)
\end{IEEEeqnarray}
\end{subequations}
where $\cs$ is defined in \eqref{eq:c2}. 
Note that $||\vect Z_{k,1}||$ follows a chi distribution of degree $\Nu$ and $F(\cdot)$ is its corresponding CDF.  By defining  $u_1 := \frac{b_1}{2}$ and $u_2 := \sqrt{\bar \gamma_{\U} + \frac{b_1^2}{4}}$ and combining this bound with the bound in \eqref{eq:52}, we prove the upper bound in \eqref{boundu}. 

%To sum up
%\begin{IEEEeqnarray}{rCl}
% \Pr [\mathcal E_{k}^{(\U)}| \mathcal E_{k,v}] \le 1- F_{||\vect Z_{k,1}||} \left (u_1 - u_2\right ) + F_{||\vect Z_{k,1}||} \left (- u_1 -u_2 \right )+ M_{\U} \lfloor 2^{\Nu R_v}\rfloor e^{-\gamma_{\U}}
%\end{IEEEeqnarray}
%---------------------------------------------------
\subsection{Bounding $\epsilon_{\e}$} 
Define the decoding error event $\mathcal E_{k,1}^{(\e)}:=\{ \hat{M}_{k}^{(\e)} \neq  \MkS\}$ for $k \in \mathcal K \backslash \Tu$ and  the decoding error event $\mathcal E_{k,2}^{(\e)}:=\{ \hat{M}_{k}^{(\e)} \neq  \MkS\}$ for $k \in \Tu$. The average error probability of decoding eMBB messages over the $K$ Tx/Rx pairs is given by 
\begin{IEEEeqnarray}{rCl}
\epsilon_{\e} = \frac{1}{K} \left ( \sum_{k \in \mathcal K \backslash \Tu} \Pr [\mathcal E_{k,1}^{(\e)}] + \sum_{k \in \Tu} \Pr [\mathcal E_{k,2}^{(\e)}]\right).
\end{IEEEeqnarray}
%----------------------
\subsubsection{Analyzing $\Pr [\mathcal E_{k,1}^{(\e)}] $ }
To evaluate this error event, we use the threshold bound for maximum-metric decoding. I.e.
\begin{IEEEeqnarray}{rCl} \label{eq:76}
 \Pr [\mathcal E_{k,1}^{(\e)}] &\le& \Pr[i_1(\vect X_{k}^{(\e,1)}, \vect X_{k}^{(\e,2)}; \vect Y_{k,1}, \vect Y_{k,2} ) \le \gamma_{\e,1}] \notag \\
&+&  M_{\e} \mathbb P[i_1(\bar {\vect X}_{k}^{(\e,1)}, \bar{\vect X}_{k}^{(\e,2)}; \vect Y_{k,1}, \vect Y_{k,2} )> \gamma_{\e,1}] 
\end{IEEEeqnarray}
for any $\gamma_{\e,1}$, where $\bar {\vect X}_{k}^{(\e,1)} \sim f_{\xkef}$ and $\bar {\vect X}_{k}^{(\e,2)} \sim f_{\xkes}$  and are   independent of $(\xkef, \xkes, \vect Y_{k,1}, \vect Y_{k,2})$.
To calculate $\mathbb P[i_1(\bar {\vect X}_{k}^{(\e,1)}, \bar{\vect X}_{k}^{(\e,2)}; \vect Y_{k,1}, \vect Y_{k,2} )> \gamma_{\e,1}]$ we follow similar steps as in \eqref{eq:44}-\eqref{eq:52} and show that 
\begin{IEEEeqnarray}{rCl} \label{eq:71}
\mathbb P[i_1(\bar {\vect X}_{k}^{(\e,1)}, \bar{\vect X}_{k}^{(\e,2)}; \vect Y_{k,1}, \vect Y_{k,2} )> \gamma_{\e,1}] \le e^{-\gamma_{\e,1}}.
\end{IEEEeqnarray}
Now to calculate $\Pr[i_1(\vect X_{k}^{(\e,1)}, \vect X_{k}^{(\e,2)}; \vect Y_{k,1}, \vect Y_{k,2} ) \le \gamma_{\e,1}]$ we first define the following distributions:
\begin{IEEEeqnarray}{rCl}
Q_{\vect Y_{k,1}}(\vect y_{k,1}) &\sim& \mathcal N (\vect y_{k,1}; 0, \syf  I_{\Nu}), \\
Q_{\vect Y_{k,2}}(\vect y_{k,2}) &\sim& \mathcal N (\vect y_{k,2}; 0, \sys  I_{\Nu}), \\
W(\vect y_{k,1}|\vect x_k^{(\e,1)}) &\sim& \mathcal N (\vect y_{k,1}; h_{k,k}\xkef , \syfx  I_{\Ne - \Nu}), \\
W(\vect y_{k,2}|\vect x_k^{(\e,2)}) &\sim& \mathcal N (\vect y_{k,2}; h_{k,k}\xkes , \sysx  I_{\Ne - \Nu}),
\end{IEEEeqnarray}
where $\syf$ and $\sys$ are defined in \eqref{eq:s2}, \eqref{eq:s3} 
and $\syfx  = 1$ and $\sysx  = 1$.
Introduce 
\begin{IEEEeqnarray}{rCl}
\lefteqn{\tilde i_1(\vect x_{k}^{(\e,1)}, \vect x_{k}^{(\e,2)}; \vect y_{k,1}, \vect y_{k,2} )} \notag \\
&:=& \ln \frac{W(\vect y_{k,1}| \vect  x_{k}^{(\e,1)} ) W (\vect y_{k,2}| \vect x_{k}^{(\e,2)})}{Q_{\vect Y_{k,1}} (\vect y_{k,1}) Q_{\vect Y_{k,2}} (\vect y_{k,2})}.
\end{IEEEeqnarray}
\begin{lemma} \label{lemma1-2}
It can be shown that 
\begin{IEEEeqnarray}{rCl}
 \frac{ i_1(\vect x_{k}^{(\e,1)}, \vect x_{k}^{(\e,2)}; \vect y_{k,1}, \vect y_{k,2} ) }{\tilde i_1(\vect x_{k}^{(\e,1)}, \vect x_{k}^{(\e,2)}; \vect y_{k,1}, \vect y_{k,2} ) } \ge J_{\e,1}
\end{IEEEeqnarray}
where
\begin{IEEEeqnarray}{rCl} \label{eq:je1}
J_{\e,1} & :=& \frac{3(\Nu-2)}{2} \ln (2) + (\Nu-2) \ln (\hkk a_1a_2)  \notag \\
&-&  \frac{3\Nu \P}{2} (\hkk^2\vnorm + a_1^2 \bef +  a_2^2 \be) \notag \\
&+&  (\Ne - \Nu - 2) \ln (2\hk \hkk) \notag \\
&-&   (\Ne - \Nu) \P(\hk^2 (1-\be) + \hkk^2  \bes)  \notag \\
&-&  \frac{e^{c_{\Gamma}} \sqrt{\P}}{\sqrt{2\pi}} \left (\frac{a_1^2 \bef }{\sqrt{\hkk^2 \vnorm}} +  \frac{ \hk^2 \be}{\sqrt{a_2^2 \be}} +  \frac{ \hkk^2 \bes}{\sqrt{\hk^2 (1-\be)}} \right) \IEEEeqnarraynumspace
\end{IEEEeqnarray}
and $a_1 := \hkk ( 1- \alpha_{k-1,1})$, $a_2 := -\hkk \alpha_{k-1,2}$ with $c_{\Gamma} \le 2$.
\end{lemma}
\begin{IEEEproof}
Similar to the proof of Lemma~\ref{lemma1}.
\end{IEEEproof}
As a result of the above lemma, we have 
\begin{IEEEeqnarray}{rCl}
  \lefteqn{  \Pr \left [i_1(\vect X_{k}^{(\e,1)}, \vect X_{k}^{(\e,2)}; \vect Y_{k,1}, \vect Y_{k,2} )\le \gamma_{\e,1} )\right]} \\
& \le & \Pr [\tilde i_1(\vect X_{k}^{(\e,1)}, \vect X_{k}^{(\e,2)}; \vect Y_{k,1}, \vect Y_{k,2} ) \le \frac{\gamma_{\e,1} }{ J_{\e,1}}] \\
& = &    \Pr  \left [\ln \frac{W (\vect Y_{k,1}| \vect  X_{k}^{(\e,1)} ) W(\vect Y_{k,2}| \vect X_{k}^{(\e,2)})}{Q_{\vect Y_{k,1}} (\vect Y_{k,1}) Q_{\vect Y_{k,2}} (\vect Y_{k,2})}  \le \frac{\gamma_{\e,1} }{J_{\e,1}} \right ] \\
& = & \Pr \Bigg [ \ln {\frac{\frac{1}{(\sqrt{2\syfx\pi})^{\Nu}}\exp \left (- \frac{|| \vect Y_{k,1} - h_{k,k}\xkef||^2}{2\syfx}\right )}{\frac{1}{(\sqrt{2\pi \syf})^{\Nu}}\exp \left (- \frac{|| \vect Y_{k,1}||^2}{2 \syf}\right )}}\notag \\
&+&  \ln {\frac{\frac{1}{(\sqrt{2\sysx\pi})^{\Ne - \Nu}}\exp \left (- \frac{|| \vect Y_{k,2} - h_{k,k}\xkes ||^2}{2\sysx}\right )}{\frac{1}{(\sqrt{2\pi \sys})^{\Ne - \Nu}}\exp \left (- \frac{|| \vect Y_{k,2}||^2}{2 \sys}\right )}} \le \frac{\gamma_{\e,1} }{J_{\e,1}}\Bigg ] \notag\\
& = & \Pr \Bigg [\frac{\syf - \syfx}{\syf\syfx}|| h_{k,k} \xkef + h_{k-1,k}( \vect X_{k-1}^{(\U)} + \vect X_{k-1}^{(\e,1)})||^2 \notag \\
&& \hspace{0.2cm}+\frac{\syf - \syfx}{\syf\syfx} || \vect Z_{k,1}||^2 +  \frac{h_{k,k}^2}{\syfx}||\xkef||^2 \notag \\
&& \hspace{0.2cm}  + 2\frac{\syf - \syfx}{\syf\syfx} \left \langle\vect Z_{k,1}, h_{k,k} \xkef  \right \rangle \notag \\
&& \hspace{0.2cm}  + 2\frac{\syf - \syfx}{\syf\syfx} \left \langle\vect Z_{k,1},  h_{k-1,k}( \vect X_{k-1}^{(\U)} + \vect X_{k-1}^{(\e,1)}) \right \rangle \notag \\
&& \hspace{0.2cm} - \frac{2h_{k,k}}{\syfx} \left \langle  h_{k,k} \xkef +  \vect Z_{k,1}, \xkef \right \rangle \notag \\
&& \hspace{0.2cm} - \frac{2h_{k,k}}{\syfx} \left \langle   h_{k-1,k}( \vect X_{k-1}^{(\U)} + \vect X_{k-1}^{(\e,1)}) , \xkef \right \rangle \notag \\
&& \hspace{0.2cm}+  \frac{\sys - \sysx}{\sys \sysx} \left( || h_{k,k} \xkes + h_{k-1,k} \vect X_{k-1}^{(\e,2)}||^2  \right )\notag \\
&& \hspace{0.2cm} + 2\frac{\sys - \sysx}{\sys \sysx} \left \langle \vect Z_{k,2},  h_{k,k} \xkes + h_{k-1,k} \vect X_{k-1}^{(\e,2)}\right \rangle \notag \\
&& \hspace{0.2cm} - \frac{2h_{k,k}}{\sysx} \left \langle h_{k,k} \xkes + h_{k-1,k} \vect X_{k-1}^{(\e,2)}+ \vect Z_{k,2}, \xkes \right \rangle \notag \\
&& \hspace{1cm} +  \frac{\sys - \sysx}{\sys \sysx}  ||\vect Z_{k,2}||^2 +  \frac{h_{k,k}^2}{\sysx}||\xkes||^2 \ge \tilde \gamma_{\e,1} \Bigg ] \IEEEeqnarraynumspace \\
& \le & \Pr \Bigg [\frac{\syf - \syfx}{\syf\syfx} \Nu \P \left (h_{k,k}^2 \be + h_{k-1,k}^2( \bu + \bef)\right ) \notag \\
&& \hspace{0.2cm} +2 \frac{\syf - \syfx}{\syf\syfx} \Nu \P h_{k,k}h_{k-1,k} \left ( \sqrt{\be \bu} + \sqrt{\be \bef}\right )  \notag \\ 
&& \hspace{0.2cm}+ 2 \frac{\syf - \syfx}{\syf\syfx} \Nu \P h_{k-1,k}^2\sqrt{\bu \bef} +\frac{\syf - \syfx}{\syf\syfx} || \vect Z_{k,1}||^2 \notag \\
&& \hspace{0.2cm} +   ||\vect Z_{k,1}|| \sqrt{\Nu \P}  \left ( 2\frac{\syf - \syfx}{\syf\syfx} \cf + \frac{2h_{k,k}}{\syfx} \sqrt{\be} \right ) \notag \\
&& \hspace{0.2cm}+  \frac{\sys - \sysx}{\sys \sysx} (\Ne - \Nu) \P \ct^2 + \frac{\sys - \sysx}{\sys \sysx} ||\vect Z_{k,2}||^2 \notag \\
&& \hspace{0.2cm} + 2||\vect Z_{k,2}|| \sqrt{(\Ne - \Nu) \P} \left (\frac{\sys - \sysx}{\sys \sysx} \ct+ \frac{h_{k,k}}{\sysx} \sqrt{1 - \be} \right )\notag \\
&& \hspace{0.2cm}+  \frac{h_{k,k}}{\sysx}\sqrt{1- \be}( \Ne - \Nu) \P \left ( \hk \sqrt{1- \be} + 2\ct \right) \notag \\
&& \hspace{1cm} +  \frac{h_{k,k}^2}{\syfx}\Nu \P \be + \frac{2h_{k,k}}{\syfx} \Nu \P   \sqrt{\be} \cf\ge \tilde \gamma_{\e,1}\Bigg ] \IEEEeqnarraynumspace \\
& = & \Pr \big  [  l_{k,1}||\vect Z_{k,1}||+  l_{k,2} ||\vect Z_{k,2}|| \notag \\
&& \hspace{0.5cm}+  l_{k,3} ||\vect Z_{k,1}||^2 + l_{k,4} ||\vect Z_{k,2}||^2 \ge \bar \gamma_{\e,1} \big ] \\
&\le & \frac{1}{\bar \gamma_{\e,1}}\mathbb E \big [  l_{k,1}||\vect Z_{k,1}||+  l_{k,2} ||\vect Z_{k,2}|| \notag \\
&& \hspace{1cm}+ l_{k,3} ||\vect Z_{k,1}||^2 + l_{k,4} ||\vect Z_{k,2}||^2 \big] \label{eq:90} \\
& = & \frac{1}{\bar \gamma_{\e,1}} \left ( l_{k,1} \sqrt{2} \frac{\Gamma(\frac{\Nu +1}{2}) }{\Gamma(\frac{\Nu}{2})} + l_{k,2} \sqrt{2} \frac{\Gamma(\frac{\Ne- \Nu +1}{2}) }{\Gamma(\frac{\Ne - \Nu}{2})} \right ) \notag \\
&& +  \frac{1}{\bar \gamma_{\e,1}} \left ( l_{k,3} \Nu + l_{k,4} (\Ne -\Nu)   \right ) \label{eq:91}
%& = & \Pr \left  [ \tilde c_1||\vect Z_{k,1}||^2 + \tilde c_2 ||\vect Z_{k,1}|| + \tilde c_3||\vect Z_{k,2}||^2 + \tilde c_4 ||\vect Z_{k,2}|| \le \bar \gamma_{\e,1} \right ]
\end{IEEEeqnarray}
where 
\begin{subequations}\label{eq:ls}
\begin{IEEEeqnarray}{rCl}
l_{k,3} & := & \frac{\syf - 1}{\syf}, \\
l_{k,4} & := & \frac{\sys - 1}{\sys}, \\
\tilde \gamma_{\e,1} & := &  -\frac{2\gamma_{\e,1} }{J_{\e,1}} + \Nu\ln \frac{\syf}{\syfx} + (\Ne - \Nu) \ln \frac{\sys}{\sysx},
\end{IEEEeqnarray}
and $l_{k,1}$, $l_{k,2}$, $\bar \gamma_{\e,1}$ are defined in \eqref{eq:lk1}, \eqref{eq:lk2}, and \eqref{eq:bgef}, respectively, $\cf$ and $\ct$ are defined in \eqref{eq:c1} and \eqref{eq:c3}. Note that in \eqref{eq:90} we employ Markov's inequality. In \eqref{eq:91}, we use the fact that $||\vect Z_{k,1}||$ follows a chi distribution of degree $\Nu$, $||\vect Z_{k,1}||^2$ follows a chi-squared distribution of degree $\Nu$, $||\vect Z_{k,2}||$ follows a chi distribution of degree $\Ne-\Nu$, and $||\vect Z_{k,2}||^2$ follows a chi-squared distribution of degree $\Ne-\Nu$ to calculate their corresponding expectations. 
\end{subequations}

%----------------------------------------------------------
%------------------------------------------------------
%------------------------------------------------------
\subsubsection{Analyzing $\Pr [\mathcal E_{k,2}^{(\e)}] $ }
To evaluate this error event, we use the threshold bound for maximum-metric decoding. I.e.
\begin{IEEEeqnarray}{rCl} \label{eq:76}
 \Pr [\mathcal E_{k,2}^{(\e)}] &\le& \Pr[i_2(\vect X_{k}^{(\e,1)}, \vect X_{k}^{(\e,2)}; \vect Y_{k,1}, \vect Y_{k,2} ) \le \gamma_{\e,2}]  \notag \\
&+&  M_{\e} \mathbb P[i_2(\bar{\vect X}_{k}^{(\e,1)}, \bar{\vect X}_{k}^{(\e,2)}; \vect Y_{k,1}, \vect Y_{k,2} )> \gamma_{\e,2}] 
\end{IEEEeqnarray}
for any $\gamma_{\e,2}$, where $\bar {\vect X}_{k}^{(\e,1)} \sim f_{\xkef}$ and $\bar {\vect X}_{k}^{(\e,2)} \sim f_{\xkes}$  and are   independent of $(\xkef, \xkes, \vect Y_{k,1}, \vect Y_{k,2})$.
To calculate $\mathbb P[i_2(\bar {\vect X}_{k}^{(\e,1)}, \bar{\vect X}_{k}^{(\e,2)}; \vect Y_{k,1}, \vect Y_{k,2} )> \gamma_{\e,2}]$ we follow similar steps as in \eqref{eq:44}-\eqref{eq:52} and show that 
\begin{IEEEeqnarray}{rCl} \label{eq:89}
\mathbb P[i_2(\bar {\vect X}_{k}^{(\e,1)}, \bar{\vect X}_{k}^{(\e,2)}; \vect Y_{k,1}, \vect Y_{k,2} )> \gamma_{\e,2}] \le e^{-\gamma_{\e,2}}.
\end{IEEEeqnarray}
To calculate $\Pr[i_2(\vect X_{k}^{(\e,1)}, \vect X_{k}^{(\e,2)}; \vect Y_{k,1}, \vect Y_{k,2} ) \le \gamma_{\e,2}]$, we first define the following distributions:
\begin{IEEEeqnarray}{rCl}
\tilde Q_{\vect Y_{k,1}}(\vect y_{k,1}) &\sim& \mathcal N(\vect y_{k,1}; \vect 0, I_n \sy), \\
\tilde Q_{\vect Y_{k,2}}(\vect y_{k,2}) &\sim& \mathcal N (\vect y_{k,2}; \vect 0, \syst  I_{\Nu}), \\
\tilde W(\vect y_{k,1}|\vect x_k^{(\e,1)}) &\sim& \mathcal N (\vect y_{k,1}; h_{k,k} ( 1- \akf) \xkef , \syfxt  I_{\Ne - \Nu}), \IEEEeqnarraynumspace\\
\tilde W(\vect y_{k,2}|\vect x_k^{(\e,2)}) &\sim& \mathcal N (\vect y_{k,2}; h_{k,k}\xkes , \sysxt I_{\Ne - \Nu}),
\end{IEEEeqnarray}
where $\sy$ and $\syst$ are defined in \eqref{eq:s1} and \eqref{eq:s4}, respectively and $\syfxt  =1$ and $\sysxt = 1$.
Introduce 
\begin{IEEEeqnarray}{rCl}
\lefteqn{\tilde i_2(\vect x_{k}^{(\e,1)}, \vect x_{k}^{(\e,2)}; \vect y_{k,1}, \vect y_{k,2} )} \notag \\
& :=& \ln \frac{\tilde W (\vect y_{k,1}| \vect  x_{k}^{(\e,1)} ) \tilde W (\vect y_{k,2}| \vect x_{k}^{(\e,2)})}{\tilde Q_{\vect Y_{k,1}}(\vect y_{k,1}) \tilde Q_{\vect Y_{k,2}} (\vect y_{k,2})}.
\end{IEEEeqnarray}
\begin{lemma} \label{lemma1-2}
It can be shown that 
\begin{IEEEeqnarray}{rCl}
 \frac{ i_2(\vect x_{k}^{(\e,1)}, \vect x_{k}^{(\e,2)}; \vect y_{k,1}, \vect y_{k,2} )  }{\tilde  i_2(\vect x_{k}^{(\e,1)}, \vect x_{k}^{(\e,2)}; \vect y_{k,1}, \vect y_{k,2} )  } \ge J_{\e,2}
\end{IEEEeqnarray}
where
\begin{IEEEeqnarray}{rCl} \label{eq:je2}
J_{\e,2} &:=& (\Nu - 2) \ln (2 \hk a_1)  + (\Ne - \Nu -2) \ln(\sqrt{2}\hkk)  \notag \\
&-&  \Nu \P (\hk^2 \vnorm + a_1^2 \be)- \frac{(\Ne - \Nu) \hkk^2 (1- \be) \P}{2}\notag \\
&-& \frac{e^{c_{\Gamma}} a_1^2 \be \P}{\sqrt{2 \pi \hk^2 \vnorm \P}} - \frac{e^{c_{\Gamma}} \hk^2 \bes \P}{\sqrt{2 \pi \hkk^2 (1- \be)\P}} - \kappa_2,
\end{IEEEeqnarray}
and $a_1 := \hkk - \hk \aks$ and $ \kappa_2 := \ln (\frac{1}{2} ) + c_{\Gamma} + \ln (\sqrt{\frac{\pi}{8}}) - 2 \ln (\hk(1-\akf))$ with $c_{\Gamma} \le 2$.
\end{lemma}
\begin{IEEEproof}
Similar to the proof of Lemma~\ref{lemma1}.
\end{IEEEproof}
As a result of the above lemma, we have
\begin{IEEEeqnarray}{rCl}
  \lefteqn{  \Pr [i_2(\vect X_{k}^{(\e,1)}, \vect X_{k}^{(\e,2)}; \vect Y_{k,1}, \vect Y_{k,2} )\le \gamma_{\e,2} )]} \\
& \le & \Pr [\tilde i_2(\vect X_{k}^{(\e,1)}, \vect X_{k}^{(\e,2)}; \vect Y_{k,1}, \vect Y_{k,2} ) \le \frac{\gamma_{\e,2} }{J_{\e,2}}] \\
& = &    \Pr  \left [\ln \frac{\tilde W (\vect Y_{k,1} |\xkef ) \tilde W (\vect Y_{k,2}| \xkes )}{\tilde Q_{\vect Y_{k,1}} (\vect Y_{k,1}) \tilde Q_{\vect Y_{k,2}} (\vect Y_{k,2})}  \le \frac{\gamma_{\e,2} }{J_{\e,2}} \right ] \\
& = & \Pr \Bigg [ \ln {\frac{\frac{1}{(\sqrt{2\syfxt\pi})^{\Nu}}\exp \left (- \frac{|| \vect Y_{k,1} - h_{k,k}\xkef||^2}{2\syfxt}\right )}{\frac{1}{(\sqrt{2\pi \sy})^{\Nu}}\exp \left (- \frac{|| \vect Y_{k,1}||^2}{2 \sy}\right )}}\notag \\
&+&  \ln {\frac{\frac{1}{(\sqrt{2\sysxt\pi})^{\Ne - \Nu}}\exp \left (- \frac{|| \vect Y_{k,2} - h_{k,k}\xkes ||^2}{2\sysxt}\right )}{\frac{1}{(\sqrt{2\pi \syst})^{\Ne - \Nu}}\exp \left (- \frac{|| \vect Y_{k,2}||^2}{2 \syst}\right )}} \le \frac{\gamma_{\e,2} }{J_{\e,2}}\Bigg ] \notag \\
& \le & \Pr \big  [  d_{k,1}||\vect Z_{k,1}|| +  d_{k,2} ||\vect Z_{k,2}||  \notag \\
&& \hspace{0.2cm}+ d_{k,3} ||\vect Z_{k,2}||^2 + d_{k,4} ||\vect Z_{k,2}||^2 \ge \bar \gamma_{\e,2} \big ] \\
&\le & \frac{1}{\bar \gamma_{\e,2}}\mathbb E \big [  d_{k,1}||\vect Z_{k,1}||+  d_{k,2} ||\vect Z_{k,2}|| \notag \\
&& \hspace{1cm}+ d_{k,3} ||\vect Z_{k,1}||^2 + d_{k,4} ||\vect Z_{k,2}||^2 \big] \label{eq:107} \\
& = & \frac{1}{\bar \gamma_{\e,2}} \left ( d_{k,1} \sqrt{2} \frac{\Gamma(\frac{\Nu +1}{2}) }{\Gamma(\frac{\Nu}{2})} + d_{k,2} \sqrt{2} \frac{\Gamma(\frac{\Ne- \Nu +1}{2}) }{\Gamma(\frac{\Ne - \Nu}{2})} \right ) \notag \\
&& +  \frac{1}{\bar \gamma_{\e,2}} \left ( d_{k,3} \Nu + d_{k,4} (\Ne -\Nu)   \right ) \label{eq:108}
\end{IEEEeqnarray}
where
\begin{subequations}\label{eq:ds}
\begin{IEEEeqnarray}{rCl}
d_{k,3} & := & \frac{\sy - 1}{\sy},  \quad
d_{k,4}  :=  \frac{\syst - 1}{\syst} 
\end{IEEEeqnarray}
and $d_{k,1}$, $d_{k,2}$, and $\bar \gamma_{\e,1}$ are defined in \eqref{eq:dk1}, \eqref{eq:dk2}, and \eqref{eq:bges}, respectively. By combining \eqref{eq:71}, \eqref{eq:91}, \eqref{eq:89} and \eqref{eq:108}, we prove the upper bound in \eqref{bounde}. 
\end{subequations}
\section{Lemmas on Bounding Distributions} \label{sec:lemmas}
\begin{lemma} \label{lemma4}
Consider the vector $\vect S = b\vect X + \vect Z$ where $||\vect X||^2 = nP$, $\vect Z  \sim \mathcal N(0, \sz I_n)$ and $b$ is a constant. Let $f_{\vect S} (\vect s)$ be the pdf of $\vect S$ and is given by 
\begin{IEEEeqnarray}{rCl}
f_{\vect S} (\vect s)
&=& \frac{1}{2(\sqrt{\sz \pi})^{n} } \Gamma(\frac{n}{2})b^{n-2} \exp (-\frac{b^2 nP}{2 \sz} ) \notag \\
&\times& \exp \left (-\frac{||\vect s||^2}{2\sz}  \right )  \frac{\mathcal I_{\frac{n}{2}-1} \left (||\vect s|| b \sqrt{nP} / \sz\right )}{\left (||\vect s|| b \sqrt{nP} / \sz\right )^{\frac{n}{2}-1}},
\end{IEEEeqnarray}
where $\mathcal I_{\frac{n}{2}-1} $ is the modified Bessel function of the first kind and $(\frac{n}{2}-1)$-th order.  
\end{lemma}
\begin{IEEEproof}
\cite[equation 52]{Molavianjazi}.
\end{IEEEproof}
\begin{lemma} \label{lemma5}
Consider the vector $\vect S = b\vect X + \vect Z$ where $||\vect X||^2 = nP$, $\vect Z  \sim \mathcal N(0, \sz I_n)$ and $b$ is a constant. Let $f_{\vect S} (\vect s)$ be the pdf of $\vect S$. Define
\begin{IEEEeqnarray}{rCl}
Q_{\vect S} (\vect s) &\sim& \mathcal N (\vect s; 0, (b^2 P  + 1)I_n), \\
\tilde Q_{\vect S} (\vect s) &\sim& \mathcal N (\vect s; 0, I_n).
\end{IEEEeqnarray}
Thus
\begin{IEEEeqnarray}{rCl}
\frac{f_{\vect S}(\vect s) }{Q_{\vect S} (\vect s)} &\le&  T_u \\
\frac{f_{\vect S}(\vect s) }{\tilde Q_{\vect S} (\vect s)} &\ge&  T_l
\end{IEEEeqnarray}
where 
\begin{IEEEeqnarray}{rCl}
T_l  &:=& 2^{\frac{n-2}{2}}b^{n-2} (e^{-b^2 P / \sz})^{\frac{n}{2}},  \label{eq:Tl}\\
T_u  &:=& e^{\kappa}, \label{eq:Tu}
\end{IEEEeqnarray}
with $\kappa := \left (\ln (\frac{1}{2} ) + c_{\Gamma} + \ln (\sqrt{\frac{\pi}{8}})  - 2 \ln (b)\right )$ and $c_\Gamma \le 2$.
\end{lemma}
\begin{IEEEproof}
Define 
\begin{IEEEeqnarray}{rCl}
T_{fq} :=  \frac{f_{\vect S}(\vect s) }{\tilde Q_{\vect S} (\vect s)} 
\end{IEEEeqnarray}
By Lemma~\ref{lemma4}, we have 
\begin{IEEEeqnarray}{rCl}
T_{fq}
&=& \frac{1}{2} \Gamma(\frac{n}{2})b^{n-2} (2 e^{-b^2 P/ \sz })^{\frac{n}{2}}  \frac{\mathcal I_{\frac{n}{2}-1} \left (||\vect s|| b \sqrt{nP} / \sz \right )}{\left (||\vect s|| b \sqrt{nP} / \sz\right )^{\frac{n}{2}-1}}. \IEEEeqnarraynumspace
\end{IEEEeqnarray}
We start by lower bounding $T_{fq}$. To this end, we use the following lower bound on the Bessel function: 
\begin{IEEEeqnarray}{rCl}
\mathcal I_n(x) > \frac{1}{\Gamma(n+1)} \left ( \frac{x}{2}\right ) ^n.
\end{IEEEeqnarray}
that is valid for $x> 0$ and $n> -\frac{1}{2}$. 
Therefore,
\begin{IEEEeqnarray}{rCl}
T_{fq}
&>&2^{\frac{n-2}{2}}b^{n-2} (e^{-b^2 P / \sz})^{\frac{n}{2}} 
\end{IEEEeqnarray}
which proves the bound \eqref{eq:Tl}. The upper bound \eqref{eq:Tu} follows the argument provided in \cite[Appendix B]{Molavianjazi}.
\end{IEEEproof}
\begin{lemma}\label{lemma6}
Consider the vector $ {\vect U} = a_1\vect X_1 + a_2 \vect X_2 + \vect Z$ where $||\vect X_1||^2 = nP_1$, $||\vect X_2||^2 = n P_2$, $\vect Z \sim \mathcal N(0, \sz I_n)$, and $a_1$ and $a_2$ are constants. Let $f_{{\vect U}} ({ \vect u})$ be the pdf of ${ \vect U}$ and 
\begin{IEEEeqnarray}{rCl}
\tilde Q_{ {\vect U}} ( {\vect u}) &\sim& \mathcal N ({\vect u}; 0, \sz I_n), \\
Q_{ {\vect U}} ( {\vect u}) &\sim& \mathcal N ({\vect u}; 0, (a_1^2P_1 + a_2^2 P_2 + \sz) I_n).
\end{IEEEeqnarray}
Thus
\begin{IEEEeqnarray}{rCl} 
\frac{f_{{\vect U}}({\vect u}) }{\tilde Q_{{\vect U}} ({\vect u})} &\ge&   (2a_1a_2)^{(n-2)} e^{-\frac{n}{2} \left ( \frac{a_1^2P_1}{\szf} + \frac{a_2^2 P_2}{\szs}\right )}, \label{eq:157} \\
\frac{f_{{\vect U}}({\vect u}) }{Q_{{\vect U}} ({\vect u})} &\le& e^{ \frac{e^{c_{\Gamma}} a_2^2P_2}{\sqrt{2\pi a_1^2P_1}}} \label{eq:158}
\end{IEEEeqnarray}
with $c_\Gamma \le 2$.
\end{lemma}
\begin{IEEEproof}
Define $\vect U_1 :=a_1 \vect X + \vect Z_1$ and $\vect U_2 := a_2\vect X_2 + \vect Z_2$ where $\vect Z_1 \sim \mathcal N(0, \szf I_n)$ and $\vect Z_2 \sim \mathcal N(0, \szs I_n)$ with $\szf + \szs = \sz$.  Let $f_{\vect U_1} (\vect u_1)$ be the pdf of $\vect U_1$, $f_{\vect U_2}(\vect u_2)$ be the pdf of $\vect U_2$, $Q_{\vect U_1} (\vect u_1) \sim \mathcal N (\vect u_1; 0,\szf I_n)$, and $Q_{\vect U_2} (\vect u_2) \sim \mathcal N (\vect u_2; 0,\szs I_n)$. Thus 
\begin{IEEEeqnarray}{rCl}
f_{{\vect U}} ({\vect u})  & = & \int_{\mathbb R ^n} f_{\vect U_1}(\vect u_1) f_{\vect U_2} ({\vect u} - \vect u_1) d \vect u_1 \\
& \ge & 2^{\frac{n-2}{2}}a_1^{n-2} (e^{-a_1^2 P_1 / \szf})^{\frac{n}{2}} \notag \\
&& \times 2^{\frac{n-2}{2}}a_2^{n-2} (e^{-a_2^2 P_2 / \szs})^{\frac{n}{2}} \notag \\
&& \times  \int_{\mathbb R ^n}   Q_{\vect U_1}(\vect u_1) Q_{\vect U_2} ({\vect u} - \vect u_1) d \vect u_1 \label{eq:164} \\
& = & (2a_1a_2)^{(n-2)} e^{-\frac{n}{2} \left ( \frac{a_1^2P_1}{\szf} + \frac{a_2^2 P_2}{\szs}\right )} \tilde Q_{{\vect U}} ( {\vect u}),
\end{IEEEeqnarray}
where the inequality \eqref{eq:164} is based on Lemma~\ref{lemma5}. This proves the lower bound in \eqref{eq:157}. The upper bound \eqref{eq:158} follows the argument provided in \cite[Appendix C]{Molavianjazi}.
\end{IEEEproof} 
%-------------------------------

\begin{lemma}\label{lemma7}
Consider the vector $\vect Y = a_1\vect X_1 + a_2 \vect X_2 +a_3 \vect X_3+  \vect Z$ where $||\vect X_i ||^2 = nP_i$ for $i \in \{1,2,3\}$, $\vect Z \sim \mathcal N(0, \sz I_n)$, and $a_i$s with $i \in \{1,2,3\}$ are constants.  Let $f_{\vect Y} (\vect Y)$ be the pdf of $\vect Y$ and 
\begin{IEEEeqnarray}{rCl}
\tilde Q_{\vect Y} (\vect y) &\sim& \mathcal N (\vect y; 0, \sz I_n), \\
Q_{\vect Y} (\vect y) &\sim& \mathcal N (\vect y; 0, (a_1 ^2 P_1 + a_ 2^2 P_2 +a_3 P_3+ \sz)I_n).
\end{IEEEeqnarray}
One can prove that
\begin{IEEEeqnarray}{rCl} 
\frac{f_{\vect Y} (\vect Y)}{\tilde Q_{\vect Y} (\vect y) } &\ge& 2^{\frac{3(n-2)}{2}}(a_1a_2a_3)^{(n-2)} \notag \\
&& \times  e^{-\frac{n}{2} \left ( \frac{a_1^2P_1}{\szf} + \frac{a_2^2 P_2}{\szs} + \frac{a_3^2 P_3}{\szt}\right )}, \label{eq:172}\\
\frac{f_{\vect Y} (\vect Y)}{ Q_{\vect Y} (\vect y) } &\le& e^{\kappa} e^{ \frac{e^{c_{\Gamma}} a_2^2P_2}{\sqrt{2\pi a_1^2P_1}}}, \label{eq:173}
\end{IEEEeqnarray}
 where $\kappa := \left (\ln (\frac{1}{2} ) + c_{\Gamma} + \ln (\sqrt{\frac{\pi}{8}}) -2 \ln (a_3)\right )$ with $c_\Gamma \le 2$, and $\szf + \szs + \szt = \sz$.
\end{lemma}
\begin{IEEEproof}
The proof  is based on the argument provided in the proof of Lemma~\ref{lemma6}. 
\end{IEEEproof}


\begin{thebibliography}{20}
\bibitem{Tataria201216g}
H.~Tataria, M.~Shafi, A.~F.~Molisch, M.~Dohler, H.~Sjöland, and F.~Tufvesson, ``6G wireless systems: vision, requirements, challenges, insights, and opportunities," \emph{Proceedings of the IEEE}, vol. 109, no. 7, pp. 1166--1199, July 2021.

\bibitem{Popovski2019}
P. Popovski, Č.~Stefanović, J.~J. Nielsen, E.~de Carvalho, M.~Angjelichinoski, K.~F.~Trillingsgaard, and A.~ Bana, ``Wireless access in ultra-reliable low-latency communication (URLLC)," \emph{IEEE Transactions on Communications}, vol. 67, no. 8, pp. 5783--5801, Aug. 2019.

\bibitem{Bairagi}
	A. K. Bairagi et al., ``Coexistence mechanism between eMBB and URLLC in 5G wireless networks,"  \emph{IEEE Transactions on Communications}, vol. 69, no. 3, pp. 1736--1749, March 2021.

\bibitem{Anand2020}
A.~Anand, G.~de Veciana, and S.~Shakkottai, ``Joint scheduling of URLLC and eMBB traffic in 5G wireless networks," \emph{IEEE/ACM Transactions on Networking}, vol. 28, no. 2, pp. 477-- 490, April 2020.

\bibitem{HomaEntropy2022}
H.~Nikbakht, M.~Wigger, M.~Egan, S.~Shamai (Shitz), J-M.~Gorce, and  H.~V.~Poor, ``An information-theoretic view of mixed-delay traffic in 5G and 6G," \emph{ Entropy}, vol.~24, no.~5, 2022. 


\bibitem{Tajer2021}
A.~Tajer, A.~Steiner, and S.~Shamai (Shitz), ``The broadcast approach in communication networks", \emph{Entropy} 2021, 23, 120.

\bibitem{shlomo2012ISIT}
	K. M.~Cohen,  A. Steiner, and S. Shamai (Shitz)
	``The broadcast approach under mixed delay constraints,'' in  \emph{Proceedings of the IEEE International Symposium on Information Theory}, Cambridge (MA), USA, July 1--6, pp.~209--213, 2012.

\bibitem{HomaITW2020}
	H.~Nikbakht, M.~Wigger, and S.~Shamai (Shitz), ``Random user activity with mixed delay traffic,'' in  \emph{Proceeding of the  IEEE Information Theory Workshop}, Apr. 11--14, 2021.

\bibitem{Cohen2022}
A.~Cohen, M.~M\'edard, and S.~Shamai (Shitz), ``Broadcast approach meets network coding for data streaming," Online: \emph{ arXiv:2202.03018v1}, Feb.~2022. 



\bibitem{ISIT2022long}
H.~Nikbakht, M.~Egan, and J-M.~Gorce, ``Dirty paper coding for consecutive messages with heterogeneous decoding deadlines in the finite blocklength regime," [Research Report] Inria - Research Centre Grenoble – Rhône-Alpes. 2022. \emph{Available: https://hal.inria.fr/hal-03556888}.

\bibitem{Costa1983}
M. H. M. Costa, ``Writing on dirty paper (Corresp.),'' \emph{IEEE Transactions on Information Theory}, vol. 29, no. 3, pp. 439–441, May 1983.

 \bibitem{Scarlett2015}
  J. Scarlett, ``On the dispersions of the Gel'fand - Pinsker channel and dirty paper coding,"  \emph{IEEE Transactions on Information Theory}, vol. 61, no. 9, pp. 4569-4586, Sept. 2015.

\bibitem{Lin2021}
	P.~H.~Lin, S. ~C.~Lin, P-W.~Chen, M.~Mross, and E.~A.~Jorswieck, ``Gaussian broadcast channels in heterogeneous blocklength constrained networks," Online: \emph{arXiv:2109.07767v1}, Sep.~2021.

\bibitem{Mross2022}
M.~Mross, P.~H.~Lin, and E.~A.~Jorswieck, ``New inner and outer bounds for 2-user Gaussian broadcast channels with heterogeneous blocklength constraints", Online: \emph{arXiv:2202.02110v1}, Feb.~2022.


\bibitem{Kassab2018}
	R.~Kassab, O.~Simeone, and P.~Popovski, ``Coexistence of URLLC and eMBB services in the C-RAN uplink: an  information-theoretic study," in  \emph{Proceeding of the IEEE Global Communications Conference},  Abu Dhabi, United Arab Emirates, Dec 9--13, 2018.

\bibitem{HomaITW2019}
	H.~Nikbakht, M.~Wigger, W.~Hachem, and S.~Shamai (Shitz), ``Mixed delay constraints on a  fading C-RAN uplink,'' in  \emph{Proceeding of the  IEEE Information Theory Workshop}, Visby, Sweden, Aug 25--28, 2019.

\bibitem{Caire2003}
G. Caire and S. Shamai, ``On the achievable throughput of a multiantenna Gaussian broadcast channel,"  \emph{IEEE Transactions on Information Theory}, vol. 49, no. 7, pp. 1691-1706, July 2003.

 \bibitem{Erseghe2016}
  T. Erseghe, ``Coding in the finite-blocklength regime: Bounds based on Laplace integrals and their asymptotic approximations," \emph{IEEE Transactions on Information Theory}, vol. 62, no. 12, pp. 6854--6883, 2016.

\bibitem{Molavianjazi}
E. MolavianJazi and J. N. Laneman, ``A second-order achievable rate region for Gaussian multi-access channels via a central limit theorem for functions,'' \emph{IEEE Transactions on Information Theory}, vol. 61, no. 12, pp. 6719--6733, Dec. 2015. 

\bibitem{Stam1982}
A.~J.~Stam, ``Limit theorems for uniform distributions on spheres in high-dimensional Euclidean spaces,” \emph{Journal of Applied Probability}, vol.~19, no.~1, pp. 221--228, 1982.

%\bibitem{Longversion}
%H.~Nikbakht, M.~Wigger, S.~Shamai (Shitz), J-M Gorce, and H.~V.~Poor, ``Joint coding of URLLC and eMBB in Wyner's soft-handoff network in the finite blocklength regime", Online: \emph{arXiv}.



 \end{thebibliography}
\end{document}